\documentclass[aps,prb,showpacs]{revtex4}
\usepackage{amsmath,amssymb}
\usepackage{graphicx}
\usepackage{dcolumn}
\usepackage{bm}
\usepackage{color}
\newcommand{\mbb}{\mathbb}
\newcommand{\mc}{\mathcal}

\newcommand{\tet}{\texttt}
\newcommand{\pr}{\partial}

\begin{document}
\title{
Peculiar electronic states, symmetries and Berry phases 
in irradiated $\alpha$-$T_3$ materials
}

\author{Andrii Iurov$^{1}\footnote{E-mail contact: aiurov@unm.edu, theorist.physics@gmail.com}$,
Godfrey Gumbs$^{2,3}$, and Danhong Huang$^{4,1}$}
\affiliation{$^{1}$Center for High Technology Materials, University of New Mexico,
1313 Goddard SE, Albuquerque, NM, 87106, USA \\
$^{2}$Department of Physics and Astronomy, Hunter College of the City
University of New York, 695 Park Avenue, New York, NY 10065, USA\\
$^{3}$Donostia International Physics Center (DIPC),
P de Manuel Lardizabal, 4, 20018 San Sebastian, Basque Country, Spain\\
$^{4}$Air Force Research Laboratory, Space Vehicles Directorate,
Kirtland Air Force Base, NM 87117, USA}

\date{\today}

\begin{abstract}
 We have laid out the results of a rigorous theoretical investigation into the 
response of electron dressed states, i.e., interacting Floquet states arising 
from the off-resonant coupling of Dirac spin-1 electrons in the $\alpha$-$T_3$
model, to external radiation with various polarizations. Specifically, we have
examined the role played by the parameter $\alpha$ that is a measure of the 
coupling strength with the additional atom at the center of the honeycomb 
graphene lattice and  which, when varied, continuously gives a different 
Berry phase. We have found that the electronic properties of the  
$\alpha$ -$T_3$ model (consisting of a flat band and two cones) could 
be  modified depending on the polarization of the imposed irradiation. We 
have demonstrated that under elliptically-polarized light the low-energy
band structure of such lattice directly depends on the valley index $\tau$. We have
obtained and analyzed the corresponding wave functions, their symmetries and the 
corresponding Berry phases, and revealed that such phases could be finite even 
for a dice lattice, which has not been  observed in the absence of the dressing 
field. This results lead to possible radiation-generated band structure engineering, 
as well as experimental and technological realization of such optoelectronic devices
and photonic crystals.  \end{abstract}

\maketitle

\section{Introduction}
\label{s1}

The $\alpha$-$T_3$ model  is considered to be the  most recent and  promising member
of novel two-dimensional (2D) materials. Their low-energy dispersions 
are obtained from a pseudospin-1 Dirac-Weyl Hamiltonian \,\cite{malcolmMain, hu} 
and possess a strong similarity with graphene.\,\cite{gr01, gr02, gr03} The 
atomic structure of the $\alpha$-$T_3$ model is represented by a honeycomb 
lattice with an additional site at the center (a hub atom) of each hexagon.
The model Hamiltonian depends on a parameter $\alpha=\tan\phi$ which 
is a measure of the coupling strength with the extra atom which depends on the 
ratio of the two hopping coefficients for all hub-rim and rim-to-rim sites.
Both  $\alpha$ and $\phi$ may be varied continuously and the Berry phase 
could be expressed in terms of these two parameters and play a crucial role 
in affecting many of the electronic and many-body properties of such 2D structures. 
 
\medskip
\par
The most encouraging technological opportunity for $\alpha$-$T_3$ is its applicability 
for tuning the value of the parameter $\phi$ from $0$ to $1$. The results for graphene 
correspond to the $\alpha\to 0$ case of this model. while $\alpha\to 1$  corresponds to 
a class of existing pseudospin-$1$ materials.\,\cite{at2, at1, malcolmMain} Such unique 
tunability together with additional electron state transitions have made studying various 
properties of $\alpha$-$T_3$ materials  as one of the most desirable directions in present-day
condensed-matter physics, chemistry and technology.

\par
\medskip 
\par

\begin{figure}
\centering
\includegraphics[width=0.49\textwidth]{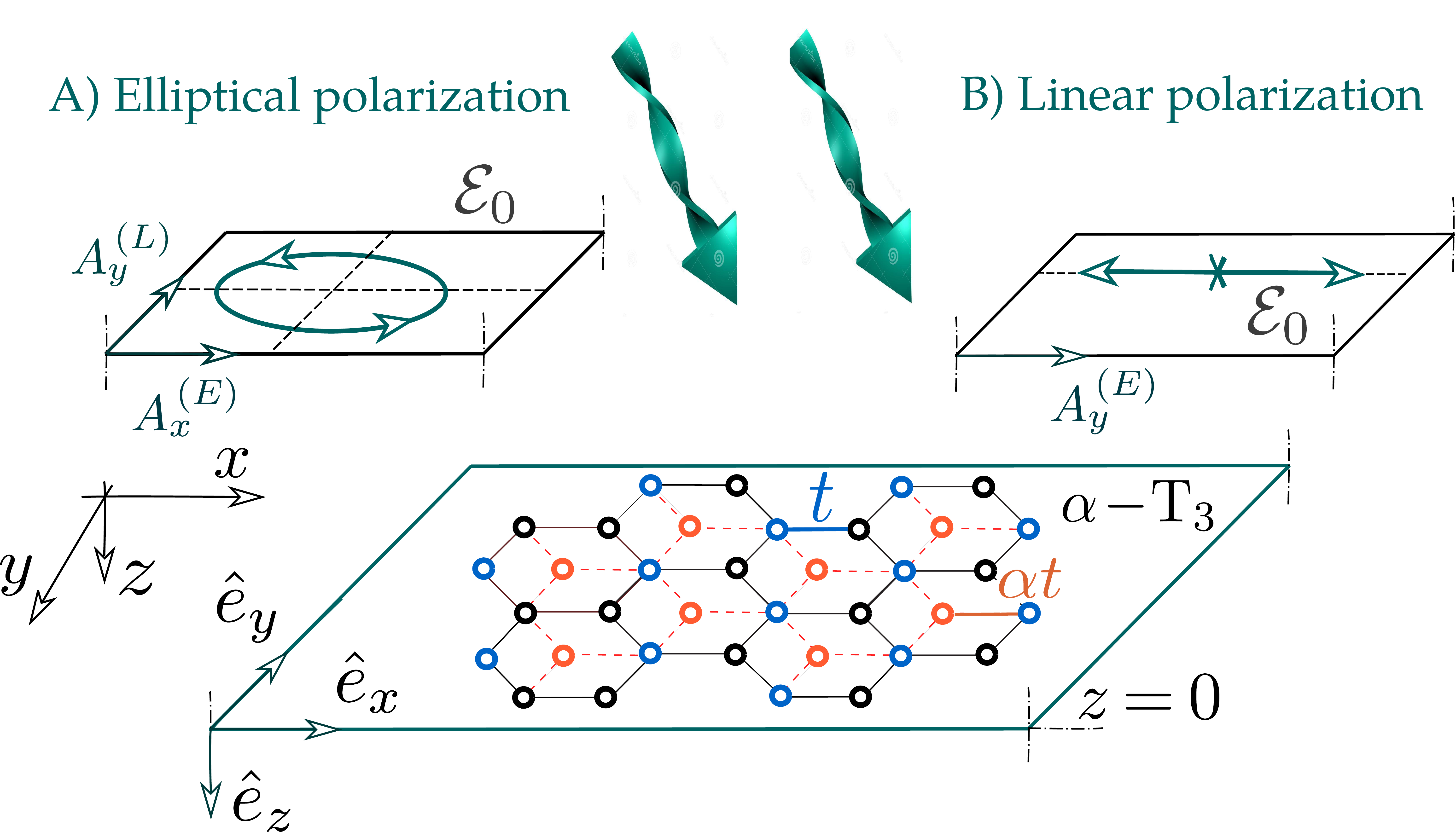}
\caption{
(Color online) Schematics of an $\alpha$-T$_3$ lattice irradiated with $A)$ 
elliptically- and $B)$ circuarly-polarized off-resonance dressing field. In each 
case, $\mc{E}_0$ is the amplitude of the electric field of the incoming wave.}
\label{FIG:1}
\end{figure}

One of the latest advances in laser and microwave technology has resulted 
in the possibility of the efficient control, as well as tunability of the basic 
electronic properties, low-energy band structure including the band gap 
and the corresponding spin- and valley-dependent electronic states by 
applying external off-resonant periodic fields. States  generated in such 
a way are referred to as \textit{electron (or optical) dressed states}
and represent a single quantum object of strongly coupled light and matter.  
They are also characterized as an electron dressed by the imposed field 
of various polarizations, just as it is schematically shown in Fig.~\ref{FIG:1}. 
The effect of such light-matter interaction on key electron properties could   
vary substantially depending on the type of polarization of the incoming 
radiation.  Most of the crucial properties of such dressed states could
be deduced from  conventional Floquet theory, which effectively describes 
an extremely wide range of quantum mechanical systems under  external 
periodic fields.\,\cite{prX1, p1, p3, p4} Based on these theories, 
researchers have developed numerous techniques to modify the existing 
electronic properties of condensed matter materials, which was 
subsequently referred to as “Floquet engineering” of various nanostructures
\,\cite{koz1, mand, Mor2} and especially for the novel low-dimensional 
Dirac cone materials.\,\cite{KiMain, 
ki1210, ourJAP2017, ki-spri, oka, dora, chak, to2} Considerable effort has 
been given to find  a way to generate topological insulator properties in 
such systems under irradiation \,\cite{Tor1,koz1,Tor2} Optical dressing
can also alter the tunneling and conductance \,\cite{ourTI} properties 
of a topological insulator, leading to tunable spin transport on their surfaces 
\,\cite{kiNJP}  with potential applications in spintronics, as
well as result in an optically-stimulated Lifshitz transition. \,\cite{ior1}

\par
\medskip 
\par 

Another important property of such electronic dressed systems for operating 
an optoelectronic device is the challenge of confining such electron states in 
a specific region. This is directly related to the presence or absence of 
the so-called Klein paradox \,\cite{Kats} in the considered material. 
Circularly-polarized radiation is known to open an energy band gap in initially 
metallic graphene \,\cite{KiMain, p2} leading to the suppression 
of  electron transmission \,\cite{ouranomalous, paula1} or electronic 
trapping \,\cite{Mor1} with a possible experimental realization. For systems with an 
already existing band gap, such as buckled honeycomb lattices, the modification 
of sich an energy gap depends on its initial value and could be either increased 
or reduced, \,\cite{ki1210} which affects all the collective electronic properties 
in a non-trivial way.\,\cite{ourPQE} In contrast, radiation with a linear 
polarization does not generate any band gaps, but leads to  strong 
anisotropy in an exposed material. \cite{kisrep}  For such anisotropic massless 
fermions, complete head-on transmission is replaced with asymmetric 
Klein tunneling \,\cite{liklein} with a crucial advance for  electronic technology, 
specifically - electron confinement in a low-dimensional material. Such Klein tunneling
in the presence of a finite $\alpha < 1$\,\cite{akl} has been shown to be different from 
both graphene and dice lattices.\,\cite{dkl}

\par
\medskip 
\par 

The principal focus  of the present work is to develop a formalism for investigating 
the properties of optical dressed states for the $\alpha$-$T_3$ model
corresponding to well-known polarizations of incoming radiations, i.e., 
elliptical (and circular as a special case) and the linear one.
Contrary to  previously published recent work,\,\cite{hu} we 
concentrate on deriving closed-form analytic approximations and the 
wave vector dependence of the obtained energy dispersions around 
each valley, determine and analyze the corresponding wave function, 
their symmetries, and the corresponding Berry phases 
which are now significantly modified by electron-light coupling. 

\par
\medskip 
\par 
The Berry phase was proven to be connected with most physical properties 
of an $\alpha$-T$_3$ lattice.\,\cite{illes01} Its orbital susceptibility undergoes 
a transition from diamagnetic in graphene to paramagnetic in a dice 
material.\,\cite{piech1, p3}, which was also shown based on a 
tight-binding model.\,\cite{piech2} The same applies to the magnetotransport of 
$\alpha$-T$_3$ materials,\,\cite{illes02} its conductivity displays
several peaks, and each of those peaks is split if $\alpha$ is set finite.\,\cite{tutul1,tutul2}
Consequently, one of our goals here will be to study how the geometric Berry phase of an $\alpha$-T$_3$
lattice is modified in the presence of a dressing field.

\par 
\medskip 
\par 

The rest of this  paper is organized in the following way. In Sec.~\ref{s2}, 
we provide a brief description of the crucial electronic properties, the low-energy 
band structure in the vicinity of the $K$ and $K^\prime$ valleys, as well 
as the corresponding electronic wave functions, noting all their possible 
relations with  phase $\phi(\alpha)$. Following this, we present our derivation 
of the off-resonant electron Floquet states for elliptically, circularly  and 
linearly polarized irradiation. Whenever possible, we also find the corresponding 
eigenstates for such dressed electrons, discuss their structure and symmetry 
properties. Specifically, we examine in which all our obtained results depend 
on $\phi(\alpha)$ and how these relations interplay with electron-field coupling. 
We also point out for which cases the obtained energy bandstructure becomes 
directly dependent on the valley index $\tau$. We also calculate Berry phases 
for all obtained previously obtained eigenstates, providing analytical 
expressions whenever possible, and discuss their features in Sec.~\ref{s3}. 
Our concluding remarks are provided in Sec.~\ref{s4} and finally, we 
provide detailed derivations of all our 
crucial results in Appendices \ref{apa}, \ref{apb} and \ref{apc}.

\section{Electron dressed states}
\label{s2}

In this section, our goal is to obtain the energy dispersions and the wave 
functions for the quasiparticle dressed states for elliptical, and circular 
as its limiting case, and linear polarizations. Considering  these situations, 
we are mainly concerned with elucidating the effect of the  phase 
$\phi(\alpha)$ on the quantities under consideration.

\par
\medskip 
\par

To establish notation, we begin with an overview of the low-energy Hamiltonian, 
its eigenfunctions and energy dispersions for the $\alpha$-$T_3$ model.
These states are derived from the $\phi$-dependent pseudospin-$1$ 
Dirac-Weyl Hamiltonian \,\cite{hu}

\begin{equation}
\label{mainH}
 \mbb{H}_\tau^{\phi}({\bf k}) = \hbar v_F \left(
  \begin{array}{ccc}
   0 & k^\tau_- \,   \cos \phi & 0 \\
    k^\tau_+ \, \cos \phi & 0 & \,  k^\tau_- \, \sin \phi   \\
   0 & k^\tau_+ \, \sin \phi  & 0
  \end{array}
 \right) \, ,
\end{equation}
where $k^\tau_\pm = \tau k_x \pm i k_y$ with $\tau=\pm$ labeling 
the valley and $v_F$ denoting the Fermi velocity. Here, $\alpha = \tan \phi$ 
represents the principal parameter, characterizing the $\alpha$-T$_3$ lattice. For 
$\alpha = 1$ and $\phi = \pi / 4$, the situation is equivalent to the Hamiltonian 
presented in Ref.~[\onlinecite{malcolmMain}]. The three solutions for the 
low-energy band structure,  $\varepsilon^{\gamma = \pm 1}_{\tau, 
\, \phi}({\bf k}) = \pm \gamma \hbar v_F k$ and 
$\varepsilon^{\gamma = \pm 1}_{\tau, \, \phi}({\bf k}) = 0$, 
introduce the valence ($ \gamma = - 1$), conduction ($\gamma 
= + 1$) and flat bands. All three energy subband energies do not  depend on the  
phase $\phi$ or $\alpha$. The corresponding wave functions are 

\begin{equation}
\label{Eig1}
 \Psi^{\gamma=\pm 1}_{\tau, \, \phi}({\bf k})  = \frac{1}{\sqrt{2}} \left(
 \begin{array}{c}
  \tau \cos \phi \,\, \tet{e}^{- i \tau \theta_{ \bf k}}  \\
  \gamma \\
  \tau \sin \phi \,\, \tet{e}^{+ i \tau \theta_{ \bf k}} 
 \end{array}
 \right) \, ,
\end{equation}
where $\theta_k = \arctan (k_y/k_x)$ is the angle associated with the wave 
vector  ${\bf k}$,   
$\varepsilon^{\gamma = \pm 1}_{\tau, \, \phi}({\bf k}) = \gamma \,\hbar v_F k$ and 

\begin{equation}
\label{Eig2}
 \Psi^{\gamma=\pm 1}_{\tau, \, \phi}({\bf k}) = \left(
 \begin{array}{c}
  \sin \phi \,\, \tet{e}^{- i \tau \theta_{\bf k}}  \\
  0 \\
  - \cos \phi \,\, \tet{e}^{+ i \tau \theta_{\bf k}} 
 \end{array}
 \right) \, ,
\end{equation}
pertaining to the flat band $\varepsilon^{\gamma = \pm 1}_{\tau, \, \phi}({\bf k}) 
= 0$. Unlike the electron dispersions $\varepsilon^{\gamma = \pm 1}_{\tau, \, \phi}({\bf k})$, most of the crucial
properties of the dressed state quasiparticle 
depend on the phase $\phi$ or on $\alpha = \tan \phi$. 

\subsection{Elliptically-polarized irradiation}
\label{s2.1}

The vector potential for elliptically-polarized light depends on the direction 
of the major axis of the polarization ellipse. Assuming that this direction is 
collinear with the $x-$axis, the expression for such a vector potential 
is\cite{ki1210}

\begin{equation}
\label{ellA}
{\bf A}^{\,(E)}(t) = \left\{
\begin{array}{c}
A^{(E)}_x (t) \\
A^{(E)}_y (t) 
\end{array}
\right\} = \frac{\mc{E}_0}{\omega} \left\{
\begin{array}{c}
\cos (\omega t) \\
\beta \, \sin (\omega t )
\end{array}
\right\} \, , 
\end{equation}
where $\beta = \sin \Theta_e$ is the ratio between the axes of the polarization 
ellipse. Equation \eqref{ellA} represents the most general case of polarization 
type, $\beta \to 1$ corresponds to circularly-polarized light with equal but phase-shifted 
components, while $\beta \to 0$ describes linearly-polarized irradiation, given by Eq.~\eqref{linA}.

\par
\medskip 
\par 

It is important to emphasize that for previously studied graphene, the 
energy dispersions obtained by semi-classical time-dependent representation 
of the vector potential \eqref{ellA} in Ref.~[\onlinecite{kisrep}] is equivalent to 
that derived using a quantized field in the limit of large photon occupation 
numbers.\,\cite{KiMain}

\par
\medskip 
\par 

Making use of the canonical substitution $k_{x,y} \rightarrow k_{x,y} - e/\hbar\,A_{x,y}$,  the initial Hamiltonian, given by Eq.\ \eqref{mainH}, is modified as 

\begin{equation}
 \label{eNH-1}
 \mbb{H}_\tau^{\phi}({\bf k}) \Longrightarrow \mc{H}^{\,(E)}({\bf k}, t) = \mbb{H}_\tau^{\phi}({\bf k}) + \hat{\mbb{H}}_A^{\,(E)}(t) \, ,
\end{equation}
where the additional interaction term is 

\begin{equation}
\label{TintHamE}
 \hat{\mbb{H}}_A^{(E)} = - \tau \, c_0 \, \left\{ 
 \,
 \left[
 \begin{array}{ccc}
  0 & \cos \phi \,\, \tet{e}^{-i \tau \, \Omega_\beta (t) } & 0 \\
  0 & 0 & \sin \phi \,\, \tet{e}^{-i \tau \, \Omega_\beta (t) } \\
  0 & 0 & 0 
 \end{array}
 \right] 
 \,\, + \,\, h.c.
 \right\} \, .
\end{equation}
Here, $\,\,+\,\,h.c.\,\,$ means adding a Hermitian conjugate matrix to the 
existing one, $\Omega_\beta (t) = \arctan \, \left[\beta \, \tan(\omega t) 
\right]$ is equivalent to $(\omega t)$ for circularly-polarized light. The interaction 
coefficient, equal to $c_0 = v_F \, e E_0/ \omega$, is equivalent to that for the 
case of linearly polarized light. $E_0$ is the amplitude of the imposed 
electromagnetic wave.  Apart from the interaction coefficient $c_0 = v_F \, e E_0/ \omega$, 
we will often use a dimensionless coupling constant 
$\lambda_0 = c_0/\hbar \omega = v_F \, e E_0/( \hbar \omega^2)$. 
Since all our studies are limited to the case of off-resonant high-frequency 
irradiation with $\hbar \omega \gg \varepsilon_d ({\bf k})$, we can always 
treat $\lambda_0 \ll a$ an infinitesimal parameter, so that the corresponding 
series expansions could be carried out. 

\begin{figure}
\centering
\includegraphics[width=0.49\textwidth]{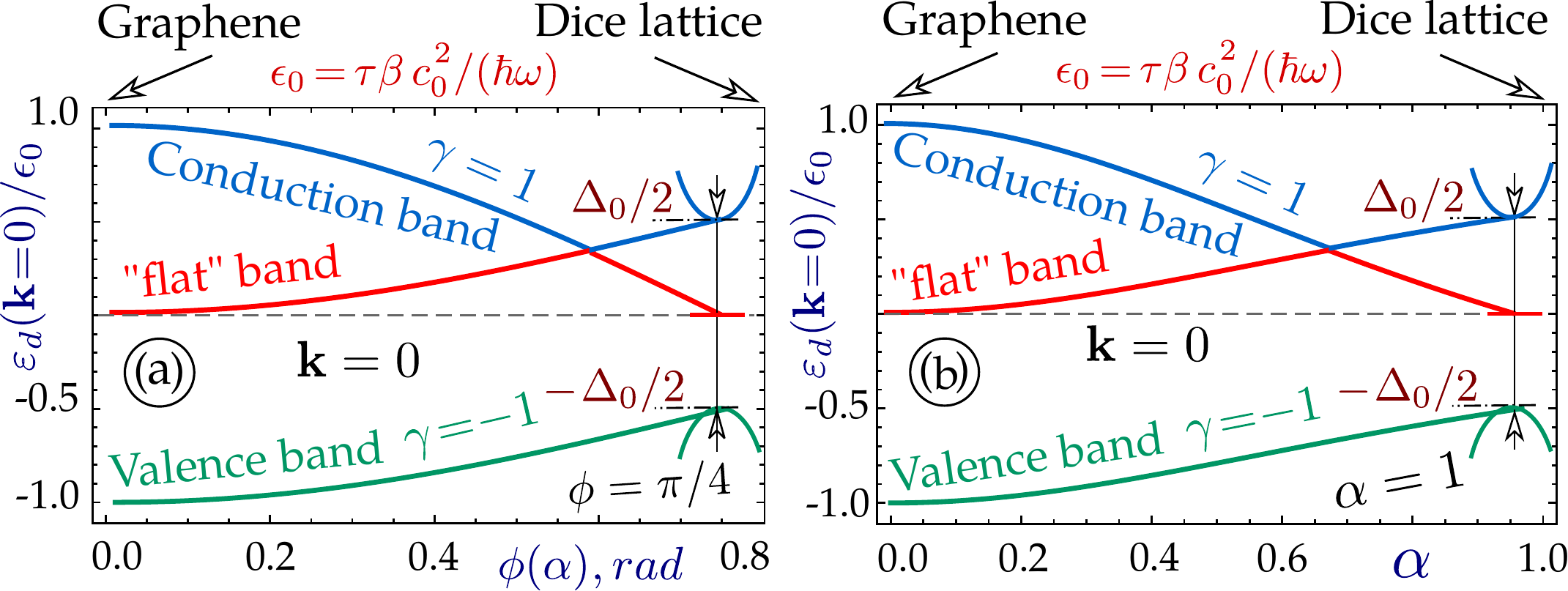}
\caption{(Color online) Zero-wavevector (measured at $K$ or $K^\prime$ point 
depending on the index $\tau = \pm 1$) energies $\varepsilon_d^{\,(e)} ({\bf k } 
= 0)$ for dressed states in $\alpha$-T$_3$ materials under an elliptically-polarized 
field as a function of parameter $\alpha = \tan \phi $ for panel $(a)$ and 
$\alpha$ for plot $(b)$. In each panel, the conduction band dispersions are 
described by a blue curve, the valence band corresponds to a green line, and 
the flat band  to the red ones. 
The energy separation between the highest (blue) and the lowest (green) 
dispersion curves defines the \textit{direct band gap} between the valence 
and conduction bands.}
\label{FIG:2}
\end{figure}
 
\par 
\medskip
\par 
 
We would now like to point out that if the initial model Hamiltonian is linear in 
$k$, which is the case for nearly all Dirac and gapped Dirac structures, such 
as gapped or gapless graphene, buckled honeycomb lattices and transition 
metal dichalcogenides, the resulting Hamiltonian for the dressed states is 
obtained by adding a single interaction term, independent of the wave 
vector ${\bf k}$. However, the situation is drastically different for  
phosphorene with a more complicated anisotropic wave vector dependence 
in the unperturbed Hamiltonian.\,\cite{ourJAP2017}

\par 
\medskip 
\par 

In order to solve this eigenvalue problem, we are going to apply \textit{perturbation
theory}.  Nearly all  off-resonant systems, subjected to external periodic fields 
with $\varepsilon_d ({\bf k}) \ll \hbar \omega$, could be effectively described by 
a perturbation expansion of the interaction Hamiltonian in  powers of $1/(\hbar \omega)$. In many cases, this would allow for an approximate solution, while 
the exact diagonalization of the interaction matrix is not possible,\,\cite{ki1210} or substantially simplify the existing calculations.\,\cite{ourJAP2017}

\par 
\medskip 
\par 

The key idea of the perturbation approach is the following. 
Once the interaction Hamiltonian term $\hat{\mbb{H}}_A^{(C)}$ is
 expressed as 

\begin{equation}
 \hat{\mbb{H}}_A^{\,(E)}  = \hat{\mbb{P}}_{\, \tau, \phi} \, \tet{e}^{i \omega t} + 
 \hat{\mbb{P}}^{\dagger}_{\, \tau, \phi} \, \tet{e}^{-i \omega t} \, , 
\end{equation}
where the operator $\hat{\mbb{P}}_{\, \tau, \phi}$ and its Hermitian 
conjugate $\hat{\mbb{P}}^{\dagger}_{\, \tau, \phi}$ are 
\textit{time-independent}, the effective Hamiltonian representing our 
dressed states becomes\,\cite{prX1}

\begin{equation}
\label{Tmexp}
 \hat{\mc{H}}_{\text{eff}}^{\,(E)} = \mbb{H}_\tau^{\phi}({\bf k}) + \frac{1}{\hbar \omega} \, \left[ 
\, \hat{\mbb{P}}_{\, \tau, \phi}, \, \hat{\mbb{P}}^{\dagger}_{\, \tau, \phi} \,
 \right] + \frac{1}{2 (\hbar \omega)^2} \left\{
 \left[ \left[
 \, \hat{\mbb{P}}_{\, \tau, \phi}, \, \mbb{H}_\tau^{\phi}({\bf k}) \, 
 \right], \, 
 \hat{\mbb{P}}^{\dagger}_{\, \tau, \phi} \,  
 \right]
 \,\, + \,\, h.c.
 \right\} \,\, + \,\, ... \, .
\end{equation}

For most of such considerations, it is sufficient to explicitly obtain two 
terms of such a power series.  In our case, the time-independent 
perturbation operator $\hat{\mbb{P}}_{\, \tau, \phi}$ is obtained as 

\begin{equation}
\label{TP}
 \hat{\mbb{P}}_{\, \tau, \phi} = - \frac{c_0}{2} \, \left[
 \begin{array}{ccc}
  0 & (\tau - \beta) \cos \phi & 0 \\
  (\tau + \beta) \cos \phi & 0 & (\tau - \beta) \sin \phi \\
  0 & (\tau + \beta) \sin \phi & 0
 \end{array}
 \right] \, .
\end{equation}
This matrix is real, but clearly not Hermitian, as it always occurs for all 
types of circularly-polarized irradiation, including the more general case 
of elliptic polarization with $0 \leq \beta < 1$.  

\par 
\medskip 
\par 

\begin{figure}
\centering
\includegraphics[width=0.50\textwidth]{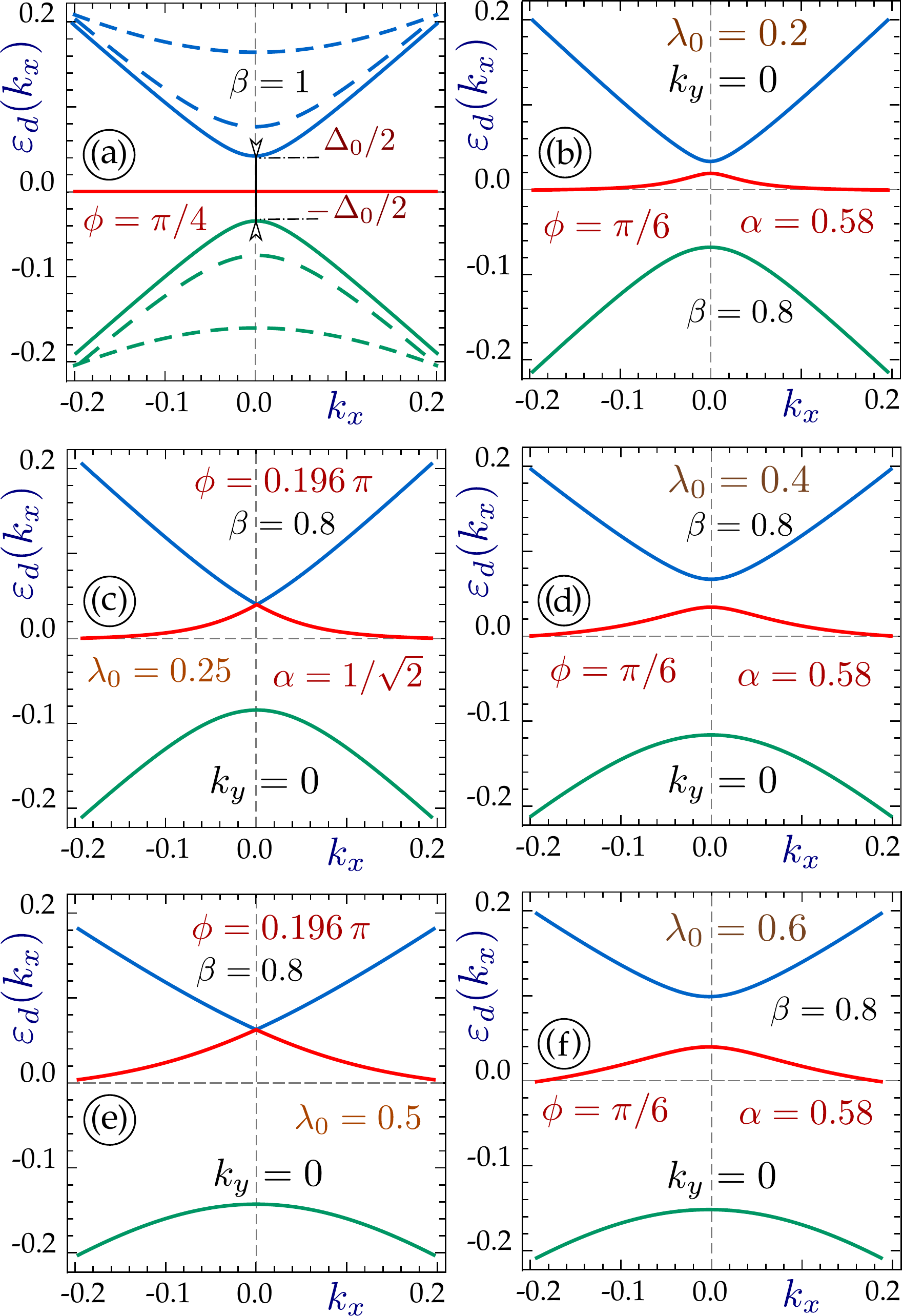}
\caption{
(Color online) Energy dispersions of a dressed state in an 
$\alpha$-$T_3$ lattice under elliptically polarized off-resonant
irradiation with $\beta = 0.8$ as a function of the $x-$component 
of the wave vector $k_x$ for $k_y = 0$.
All results are obtained 
in the vicinity of the $K$ valley with $\tau = 1$.
Each plot represents the dispersions corresponding to three subbands,
i.e., the conduction ($\gamma = 1$, upper blue curve), valence 
($\gamma = - 1$, lower green line) and the ``flat'' band (middle  red curve).
Panel $(a)$ demonstrates the energy dispersions for a dice lattice ($\phi = \pi/4$). 
All the remaining left plots $(c)$ and $(e)$ correspond to the phase $\phi=0.196\pi$
or $\alpha = 1/\sqrt{2}$, for which one of the band gaps is zero.
 All the right panels 
$(b)$, $(d)$ and $(f)$ are related to $\phi=\pi/6$ and are distinguished by the value 
of the dimensionless electron-light coupling constant $\lambda_0 = 0.2$ for plot $(b)$, 
$0.4$ for $(d)$ and $0.6$ for $(f)$, similar to three different curves for each 
band in plot $(a)$. Panel $(c)$ corresponds to $\lambda_0 = 0.25$, and plot $(e)$  
to 0.5.
}
\label{FIG:3}
\end{figure}

 \par
\medskip
\par
Evaluating the commutation relation in Eq.~\eqref{Tmexp}, we arrive at the 
following expression for the effective perturbation Hamiltonian

\begin{equation}
\label{TeffH}
\hat{\mc{H}}_{\text{eff}}^{\,(E)} ({\bf k},t)= \mbb{H}_\tau^{\phi}({\bf k}) -
\tau \beta \, \lambda_0 \, c_0 \, \left[
 \begin{array}{ccc}
  \cos^2 \phi & 0 & 0 \\
  0 & -\cos(2 \phi) & 0 \\
  0 & 0 & -\sin^2 \phi
 \end{array}
 \right] - \frac{\tau}{4} \,  \hbar v_F \, \lambda_0^2 \, 
\left[
 \begin{array}{ccc}
  0 & h_{12} & 0 \\
    h_{12}^{\star} & 0 & h_{23} \\
  0 &  h_{23}^{\star} & 0
 \end{array}
 \right] \, ,
\end{equation}
where $h_{12}(\beta \, \vert \, \tau, \phi) =  \cos \phi \left[ 1 + 3 \cos (2 \phi) \right] \,(\beta^2 \, kx - i \tau k_y)$
and $h_{23}(\beta \, \vert \, \tau, \phi) = \sin \phi \left[ 1 - 3 \cos (2 \phi) \right] \, (\beta^2 \, kx - i \tau k_y)$.

\par 
\medskip
\par

It is interesting to note that the chirality $\tau k_x \pm i k_y$ of 
the second perturbation term is the same as that in the unperturbed 
Hamiltonian $\mbb{H}_\tau^{\phi}({\bf k})$ given by 
Eq.~\eqref{mainH}. 

\par 
\medskip
\par
One needs to pay close attention to the first perturbation term, which is 
independent of the wave vector ${\bf k}$ and determines the position 
of each band edge at the $K$ point

\begin{equation}
  \varepsilon_d^{\,(E)} ({\bf k} = 0) = \tau \beta \frac{c_0^2}{\hbar \omega} \, \times \,
   \begin{cases}
       - \cos^2 \phi \\
       + \cos (2 \phi) \\
       + \sin^2 \phi 
                  \end{cases} = \tau \beta \, \lambda_0 c_0 \, \times \, 
                  \begin{cases}
                   - \left(1 + \alpha^2\right)^{-1} \\
                   \left(1 - \alpha^2\right) / \left(1 + \alpha^2 \right) \\
                  \alpha^2 / \left( 1 + \alpha^2 \right) \, .
                  \end{cases}
\end{equation}
as well as the energy gap between the valence and conduction bands

\begin{equation}
 \label{gap0}
 \Delta_0 (\beta, \phi \, \vert \, \lambda_0) = 
 \frac{\beta}{2} \, \lambda_0 \,c_0  \begin{cases} 
                                                 \cos (2 \phi) + \cos^2 \phi \,\,\,\, \text{for} \,\,\,\,
                                                 0 \leq \phi < 0.615\,rad \, ,  \\
                                                 {\bf 1.0} \hskip0.86in \text{for} \,\,\,\,
                                                 0.615\,rad < \phi \leq \pi/4  \, . 
                                                \end{cases}
\end{equation}
These simple but important results for the zero-momentum band structure 
are displayed in Fig.~\ref{FIG:2}.  Each energy is given in  units of 
$\epsilon_0 = \tau \beta \, c_0^2/(\hbar \omega)$, i.e., its actual 
value depends on the valley index $\tau$, the ratio of the 
ellipse polarization  axes $\beta$ and the electron-light interaction strength. 
The ``flat" band  in fact no longer posses this property, all  three 
subbands are now distorted and intersect with each other. The energy 
band gap differs from $\lambda_0 \, c_0$ for graphene to 
$\lambda_0 \, c_0 / 2$ for a dice lattices. It is interesting to see 
that the band gap does not change for $\phi > 0.615\,rad \backsimeq 0.196 \, \pi$ or 
$\alpha > 1/\sqrt{2}$. The band gap for a dice lattice is the smallest and is
exactly half of that for graphene (the largest). The $\alpha$-dependence 
of the band edge locations, presented in Fig.~\ref{FIG:2} $(b)$ for 
comparison, is similar but not exactly identical with the corresponding 
dependence on $\phi$.

\par 
\medskip 
\par 

An analytical solutions for finite-$k$ energy dispersion could in principle be 
obtained, as it is always true for a third-power algebraic equation.  
However, the obtained expressions are quite lengthy and complicated so that they 
cannot be conveniently analyzed. 

\par
\medskip 
\par

Our numerical results for the energy dispersions for the case of elliptically-
polarized light are presented in Fig.~\ref{FIG:3}. Leaving out the simplest 
case of $\phi=\pi/4$, which was discussed above and presented in 
Fig.~\ref{FIG:3} $(a)$, we examine the remaining cases and 
see that the initially flat $\varepsilon_d ({\bf k}) = 0$ subband 
acquires a ${\bf k}$-dependent non-zero curvature and may be 
located either above or below the $\varepsilon = 0$ level 
depending on the phase as well as the valley index $\tau$. 
The valence and conduction band edges are shifted in  
the vertical direction, so that there is no longer symmetry 
between the valence and conduction bands, as it was also 
pointed out in Ref.~[\onlinecite{hu}]. At the same 
 time, there is complete inversion symmetry for
 $k_{\{x,y\}} \to - k_{\{x,y\}}$, which means that only 
even powers of the wave vector components  $k_x$ and $k_y$ 
are present in the general eigenvalue equation based on the 
effective perturbation Hamiltonian \eqref{TeffH}. For all cases, 
the effect of the imposed irradiation on the
flat band and its displacement from the zero-energy level
are most noticeable at ${\bf k} = 0$. We emphasize that the  
anisotropy and the angular dependence in these dispersions 
appear only because of the anisotropy of the external perturbation 
field and is completely absent for the case of circularly-polarized 
irradiation with $\beta = 1$.

\begin{figure}
\centering
\includegraphics[width=0.50\textwidth]{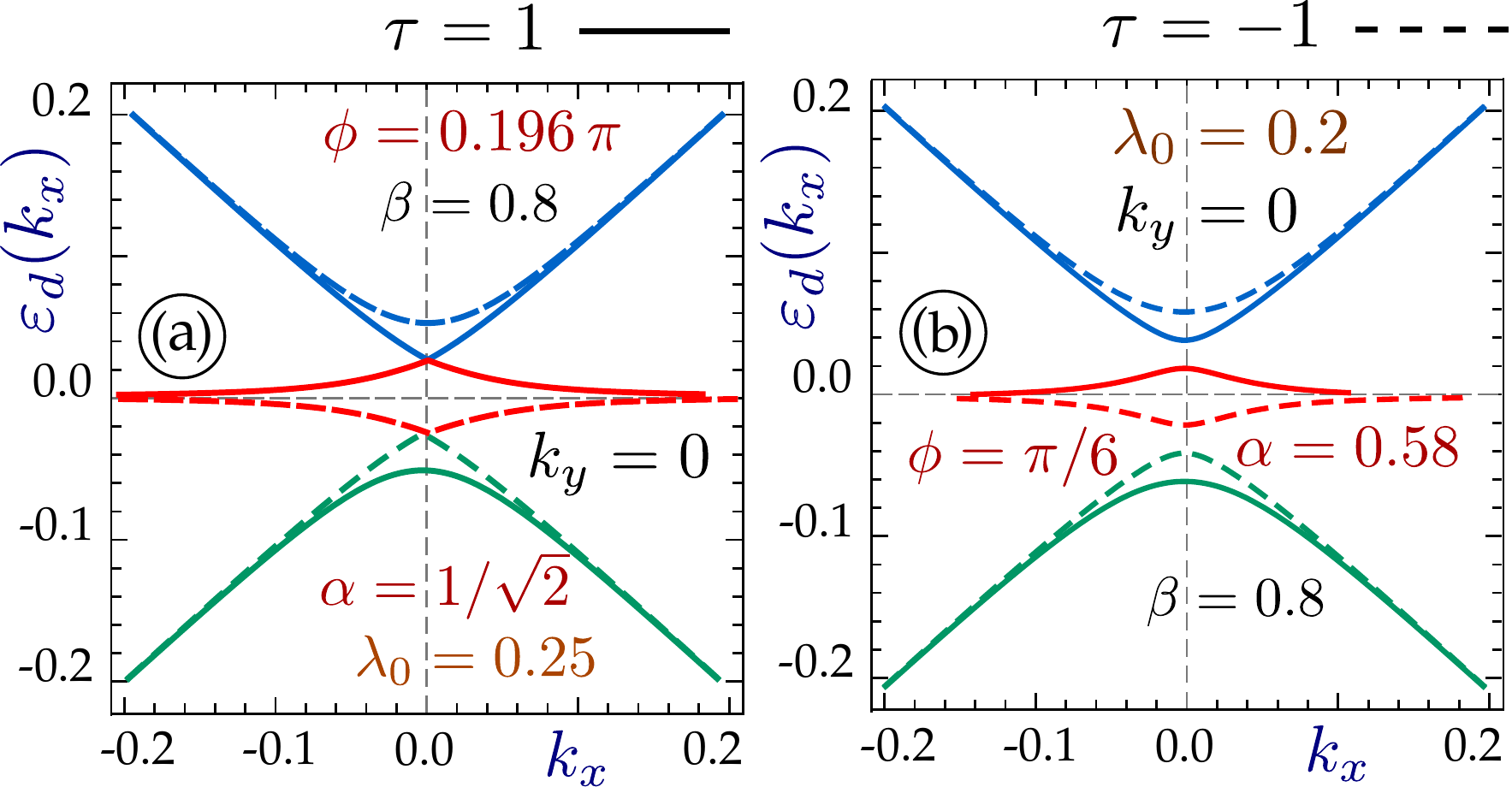}
\caption{(Color online) Effect of valley index $\tau$ on the 
energy dispersions of a dressed state in an $\alpha$-$T_3$ 
lattice under elliptically-polarized dressing field. We plot the 
energy dispersions $\varepsilon_d^{\,(E)}({\bf k})$ as a 
function of the $x-$component of the wave vector $k_x$ for 
$k_y = 0$. Panel $(a)$ describes the dispersions for 
$\alpha = 1\sqrt{2}$ when one of the gaps is closed, while panel 
$(b)$ shows plots for $\phi = \pi/6$ and $\alpha = 0.586$.
In each plot, the solid lines correspond to the $K$ valley or 
$\tau = + 1$, and the dashed ones to the $K^\prime$  
with $\tau = - 1$ and the dimensionless electron-light 
coupling constant $\lambda_0 = 0.25$. }
\label{FIG:4}
\end{figure}

Not only the energy band gaps and edges from Eq.~\eqref{gap0}, 
but also the whole finite-$k$ energy dispersion branches demonstrate 
direct and substantial dependence on the valley index $\tau = \pm 1$, 
as we see from Fig.~\ref{FIG:4}. The location  of each energy subband
 varies near the $K$ and $K^\prime$ valleys, and this noticeable 
difference is not limited to a pure change of $\pm$ sign, in contrast to 
all situations observed before. For $\alpha = 1/\sqrt{2}$, either upper 
or lower bandgap is closed, depending on the value of $\tau$. As a 
result, we obtain non-equivalent density of electronic states in each 
low-energy ($K$ or $K^\prime$) region. The contributions from such 
electronic states from each valley is no longer equal or opposite to 
each other, which could strongly affect the transport properties. 

\par 
\medskip
\par
These obtained dispersions show a striking resemblance to
silicene, in which the electronic states with a given spin 
$\sigma = \pm 1$ demonstrate two inequivalent band gaps 
$\Delta_{\tau,\sigma} = \vert \Delta_{SO} - \sigma \tau \Delta_z \vert$, 
in which $\Delta_{SO}$ is a constant intrinsic spin-orbit gap and 
$\Delta_z$ can accept nearly any values depending on the external 
perpendicular electrostatic field. Consequently, varying the applied field, the 
gap between the bands can be opened or closed and a buckled honeycomb 
lattice appears to be in a topological insulator, valley-spin polarized metal 
(not gap) or a conventional band insulator state. The lower gap 
$\vert \Delta_{SO} - \Delta_z \vert$, which defines the actual band 
gap between the valence and conduction bands, becomes equal to the 
upper one $\Delta_{SO} + \Delta_z$, once either valley index $\tau$ 
or spin index $\sigma$ is changed to its opposite 
value\,\cite{SilMain,EzawaSi}. This gives rise to specific transport 
properties\,\cite{ournewprb} and a number of valleytronics 
applications,\,\cite{hua10} which could now also be based on the 
irradiated $\alpha$-T$_3$ lattice. 

\par 
\medskip 
\par 

Finally, for graphene with $\varepsilon (\vert {\bf k} \vert) = 
\pm \hbar v_F \vert { \bf k } \vert$,  
interacting with circularly-polarized irradiation,
the off-resonant dressed states have the following dispersion relations  

\begin{equation}
 \varepsilon_d^{\gamma= \pm 1} ({\bf k}) = \pm \left\{  
  \left(\frac{c_0^2}{\hbar \omega}\right)^2 + \left[ \,
 \hbar v_F k \, 
 \left\{ 
 1 - 2 \,\left( \frac{c_0}{\hbar \omega} \right)^2
  \, 
  \right\} \right]^2 \,
 \right\}^{1/2} \, , 
\end{equation}
which could be obtained as a limiting case of vanishing anisotropy for 
multi-layer black phosphorus, \,\cite{ourJAP2017} or setting to zero 
all the band gaps for transition  metal dichalcogenides which 
were investigated in Ref.~[\onlinecite{ki1210}]. This result for graphene 
is a $\backsimeq 1/(\hbar \omega)^2$ approximation of well-known 
circularly-polarized irradiated quasiparticle dispersions. In contrast, 
an exact solution shows \,\cite{KiMain} that the energy band gap is 
equal to 

\begin{equation}
 2 \Delta_0(c_0 \, \vert \, \omega) = \sqrt{(\hbar \omega)^2 +
4 \, c_0^2} -\hbar \omega \backsimeq \frac{2 \, c_0^2}{\hbar \omega} \cdot \left[ 1 - \left(
\frac{c_0}{\hbar \omega}  \right)^2 + \,\,...\,\,
\right] \, , 
\end{equation}
and the Fermi velocities in each direction are not affected. We again see 
that the field-induced energy bandgap for graphene is exactly twice as 
large when compared to a dice lattice. 

\subsection{Symmetric band structure and wave function for a dice lattice}
\label{s2.2}

Here, we address a specific case of spin-1 dice lattices without any 
dependence on $\alpha$. This corresponds to $\phi=\pi/4$.
Most importantly, all the equations are greatly simplified and additional 
closed-form analytical results could be obtained and investigated. 
Also it is crucial to see that the effect of irradiation is the lowest in this case 
for a given dressing field intensity, as we see from Figs.~\ref{FIG:3}$(c)$ and \ref{FIG:3}$(d)$. 
This case has special significance for technical applications 
since those T$-3$ spin-1 materials could be possibly fabricated at the present time. 

\par 
\medskip
\par 
In this case, the non-interacting Hamiltonian \eqref{mainH} for a dice lattice 
takes the form 

\begin{equation}
 \mbb{H}_\tau^{d}({\bf k}) = \frac{\hbar v_F}{\sqrt{2}} \, \left[
 \begin{array}{ccc}
  0 & k_\tau^{-} & 0 \\
  k_\tau^{+} & 0 & k_\tau^{-} \\
  0 & k_\tau^{+} & 0
 \end{array}
 \right] = 
 \sum \limits_{\alpha = \pm} \hat{\Sigma}_\alpha^{\,(1)} \, k_\tau^{\, \alpha} \, ,
\end{equation}
where $\hat{\Sigma}_{\pm 1}^{\,(1)} = \hat{\Sigma}_{x}^{\,(1)} \pm i \hat{\Sigma}_{y}^{\,(1)}$ are 
defined based on spin-1 matrices in Appendix \ref{apa}. 

\par 
\medskip

The effective perturbation Hamiltonian \eqref{Tmexp} in this case up 
to the order of $1/(\hbar \omega)^2$ is  

\begin{equation}
\label{H3}
 \hat{\mc{H}}_{\text{eff}}^{\,(E)} ({\bf k},t) \backsimeq \left[
 \begin{array}{ccc}
   - \tau \beta/4 \cdot \lambda_0 \,c_0 & \mc{O}_{\,\tau}^{\,(-)}(\beta \, \vert \, \lambda_0, {\bf k})
   & 0 \\
  0 & 0 & \mc{O}_{\,\tau}^{\,(-)}(\beta \, \vert \, \lambda_0, {\bf k}) \\
  0 & 0 & \tau \beta/4 \cdot \lambda_0 \,c_0
 \end{array}
 \right] \,\, + \,\, h.\,c. \, ,
\end{equation}
where 
\begin{eqnarray}
&& \mc{O}_{\,\tau}^{\,(-)}(\beta \, \vert \, \lambda_0, {\bf k}) =  \frac{ \hbar v_F}{\sqrt{2}} \, \left[ 
   \tau \, k_x \left(1 + \beta^2 \, \frac{\lambda_0^2}{4} \right) - i k_y \left(1 + \frac{\lambda_0^2}{4}
   \right)
   \right] \, .
\end{eqnarray}
Such a dramatic simplification of Eq.~\eqref{TeffH} has been made possible 
mainly due to the fact that 
in our case \(h_{12}(\beta \, \vert \, \tau, \phi) = h_{23} (\beta \, \vert \, \tau, \phi = \pi/4)= 
1/\sqrt{2} \cdot (\beta^2 \, kx - i \tau k_y)\).

\par 
\medskip
\par 
For a Hamiltonian with such symmetry, the low-energy band structure is 
symmetric for electrons and holes with 

\begin{eqnarray}
\label{SpecRes}
&& \varepsilon_d^{\,(E)} (\beta \, \vert \, \lambda_0, {\bf k}) = 0 \, \hspace{0.1in} \, \text{and} \, \hspace{0.1in} 
 \varepsilon_d^{\,(E)} (\beta \, \vert \, \lambda_0, {\bf k}) = \pm \sqrt{\mc{S} \left(
 \beta \, \vert \, \lambda_0, {\bf k}
 \right) } \, ,
 \hspace{0.1in} \text{where} \\
 \nonumber 
&& \mc{S}\left(\beta \, \vert \, \lambda_0, {\bf k} \right) = 
\left( \frac{\beta}{4} \, \lambda_0 \, c_0 \right)^2 + \left( \hbar v_F \right)^2 \left\{ k^2 + 
\frac{\lambda_0^2}{2} \left[ 
(\beta k_x)^2 \,\left(
1 + \frac {\beta^2 \, \lambda_0^2}{8} 
\right) + k_y^2 \, \left(
1 + \frac {\lambda_0^2}{8} 
\right) \,
\right] \, 
\right\}\, .
 \end{eqnarray}
Only in this case, do the middle subbands remain flat for all wave vectors and 
the valence and conduction subbands stay completely symmetric for 
$\gamma = \pm 1$.

\par 
\medskip
\par 

Equation~\eqref{gap0} shows the opening of a finite energy band gap twice 
as large as $\Delta_0(\beta, \lambda_0) = \beta \lambda_0 \, c_0/2$. It is 
clear that the anisotropy of the energy bands is caused solely by the elliptical
polarization of the dressing field and disappears for $\beta = 1$. From now on, 
we will focus only on the latter case of circularly-polarized irradiation. In this way, 
the energy label $\gamma = \pm 1$ still represents a Dirac cone with 
\textit{renormalized isotropic} Fermi velocity $v_F^{\,(r)}(\lambda_0) =
v_F \left\{ 1 + \lambda_0^2/2 \left(1 + \lambda_0^2/8 \right) \right\}$.

\par 
\medskip 
\par 
We introduce the following simplifying notation: \(\delta(\lambda_0,  k) = 
\Delta_0(\beta=1, \lambda_0)/\left[ \hbar \, v_F^{\,(r)}(\lambda_0) 
\, k \right]\). This quantity is dimensionless, and it represents the induced 
energy gap related to a fixed electron energy for chosen non-zero wave vector 
$k$. For our case of small electron-light coupling with $\lambda_0 \ll 1$, 
it could be approximated by

\begin{equation}
 \delta(\lambda_0 \ll 1,  k) \backsimeq \frac{\lambda_0}{2} \, \frac{c_0}{\hbar v_F k} \, \left(
 1 - \frac{\lambda_0^2}{2} \,
 \right) \, .
\end{equation}

\par 
\medskip 
\par 

Now, we can obtain the wave functions, corresponding to dispersions 
in \eqref{SpecRes} as the eigenstates of the matrix \eqref{H3}. For the sake 
of simplicity, we will consider circularly-polarized light with $\beta = 1$.
The solutions, pertaining to the valence and conduction bands 
$\gamma = \pm 1$ are 

\begin{equation}
 \label{Twf1}
 \Psi_d^{\gamma = \pm 1}(\tau\, \vert \, \lambda_0, {\bf k}) = 
 \frac{1}{\sqrt{\mc{N}}} \, \left\{ 
 \begin{array}{c} 
  \tau \, \mc{C}^{(1)} \, \tet{e}^{- i \tau \theta_{\bf k}}  \\
  \mc{C}^{(2)} \\
   \tau \, \tet{e}^{+ i \tau \theta_{\bf k}}  
 \end{array}
 \right\} \, .
\end{equation}
where the component amplitudes and normalization factor may be obtained.
In our case of low intensity of the incident radiation $c_0 \ll \hbar \omega$ 
these expressions could be  simplified as 

\begin{eqnarray}
 && \mc{C}^{(1)}(\gamma \, \vert \,\lambda_0 \ll 1, k) 
 \backsimeq 1 + 2 \, \delta(\lambda_0, k) \left[
 \gamma + \delta(\lambda_0, k) \, \right]
 \, ,  \\
 \nonumber 
 && \mc{C}^{(2)}(\gamma \, \vert \, \lambda_0 \ll 1, k) \backsimeq 
 \sqrt{2} \, \left[ \gamma - \delta(\lambda_0, k) \, \right] +
 \frac{\gamma}{\sqrt{2}} \,  \delta^2(\lambda_0, k) 
 \, ,\\
 \nonumber 
 && \mc{N}(\gamma \, \vert \, \lambda_0 \ll 1, k) \backsimeq 4 \left[
 1 + 3 \, \delta^2(\lambda_0, k) \,
 \right] \, .
\end{eqnarray}

\par 
\medskip 
\par

The components of the wave function \eqref{wf01} are no longer equal to 
each other (apart from a common phase difference), as it is expected to occur 
when an energy gap is opened. Comparable alteration of the wave function components
was demonstrated for irradiated graphene.\,\cite{ouranomalous} 
These components now also depend on the electron-hole index 
$\gamma = \pm 1$, i.e., their deflection from their initial $1/\sqrt{2}$ value 
is not the same for  electrons and holes.  

\par 
\medskip 
\par

The remaining wave function for the flat band ($\gamma = 0$) is 

 \begin{equation}
\label{Twf2}
\Psi_d^{\gamma = 0}(\tau\, \vert \, \lambda_0, k) = \frac{1}{ \sqrt{
  2 \left[
 1 + \delta^2(\lambda_0, k)
 \, \right]
 }} \,
 \left\{  
 \begin{array}{c}
  \tet{e}^{- i \tau \theta_{ \bf k}}  \\
  \sqrt{2}\,\delta(\lambda_0,  k) \\
  - \tet{e}^{+ i \tau \theta_{ \bf k}} 
 \end{array}
 \right\} \, .
\end{equation}
Here, one of its components is also inequivalent to the two others. This field-induced 
modification of the middle component and the normalization of the wave function \eqref{Twf2} does not depend on the valley index $\tau$. 

 \subsection{Linear polarization of the incoming radiation}
\label{s2.3}

We now turn our attention to an alternative situation  in which linearly-polarized 
radiation is incorporated into  the $\alpha$-$T_3$ model Hamiltonian. 
Being essentially anisotropic, such fields are  known to transform the Dirac 
cone into an asymmetric elliptical cone without creating a gap between the 
valence and conduction bands.\cite{kisrep} Whereas in initially anisotropic 
phosphorene, the  direction of the linear polarization was important,\cite{ourJAP2017}
for the isotropic energy dispersions in $\alpha-T_3$ we can assume
that the polarization vector lies along the $x$ axis 
without any loss of generality so that 

\begin{equation}
\label{linA}
{\bf A}^{(L)}(t) = 
\left\{ \begin{array}{c}
          A^{(L)}_x (t) \\
          0
        \end{array}
\right\} = \frac{\mc{E}_0}{\omega} \left\{ 
\begin{array}{c}
\cos (\omega t )\\
0
\end{array}
\right\} \, . 
\end{equation}

\par 
\medskip 
\par
In each case of linear k-depndence, the total Hamiltonian of an interacting 
quasiparticle only acquires an additional term given by

\begin{equation}
\label{Tlinham}
 \mbb{H}_\tau^{\phi}({\bf k}) \Longrightarrow \hat{\mc{H}}^{(L)}({\bf k}, t) = 
 \mbb{H}_\tau^{\phi}({\bf k})+\hat{\mbb{H}}_A^{(L)} \, , 
\end{equation}
where 

\begin{equation}
 \label{HAL}
 \hat{\mbb{H}}_A^{(L)} = - \tau \, c_0 \cos (\omega t) \, \left[
 \begin{array}{ccc}
  0 & \cos \phi & 0 \\
  \cos \phi  & 0 & \sin \phi \\
  0 & \sin \phi & 0 
 \end{array}
 \right] \, .
\end{equation}
The coupling constant $c_0 = v_F e E_0/\omega$ is equivalent to that in the 
case of elliptically- or circularly-polarized light. We also note that each element 
of the matrix in \eqref{HAL} has identical periodic time dependence, which was 
not true for circularly-polarized irradiation field. 

\par 
\medskip 
\par

The case of linearly-polarized dressing field is distinguished because the 
time-dependent Schr{\H o}dinger equation at $K$ (or $K^\prime$) point 
for ${\bf k}=0$ \textit{could be solved exactly}. This means that our result
regarding the absence of an energy bandgap is precise and, most crucially, 
a wave function with appropriate time dependence could be obtained in 
contrast to the previous case of elliptically polarized light. 

\par 
\medskip 
\par 

The detailed derivation of the energy dispersions and the wave function 
for the linearly-polarized dressing field is provided in Appendix \ref{apb}. 
The dressed state quasiparticle energy dispersions are given by 

\begin{eqnarray}
\label{linD}
 && \varepsilon_d^{\,(L)} (\phi \, \vert \, \lambda_0, {\bf k}) = 0 \,\,\, \text{and} \\
 \nonumber 
 && \varepsilon_d^{\,(L)} (\phi \, \vert \, \lambda_0, {\bf k}) = \pm \hbar v_F  \, 
 \sqrt{\mbb{A}(\theta_{\bf k} \, \vert \, \phi, \lambda_0)} \, k \, ,
\end{eqnarray}
with angular dependence 

\begin{equation}
\label{TA}
\mbb{A}(\theta_{\bf k} \, \vert \, \phi, \lambda_0) = 
 \cos^2 \theta + \left\{
 \left[
 \mc{J}_0 (2 \lambda_0) \, \cos (2 \phi)
 \right]^2 + 
 \left[
 \mc{J}_0 (\lambda_0) \, \sin (2 \phi)
 \right]^2
 \right\} \,\, \sin^2 \theta
\end{equation}
clearly indicates  \textit{field-induced anisotropy}. The corresponding 
dispersion relations for graphene, obtained in Ref.~[\onlinecite{kisrep}], 
are immediately recovered if $\phi$  or $\alpha$ is set equal to zero. In 
the opposite limit for a \textit{dice lattice} $\phi = \pi/4$,  only the second
term $\mc{J}_0 (\lambda_0) \, \sin (2 \phi)$ in Eq.~\eqref{TA} remains 
finite, so that the effect of electron-field interaction is the lowest.

 \par 
\medskip 
\par

Since our consideration is executed for off-resonant radiation with 
$\lambda_0 = c_0/ (\hbar \omega) \ll 1$, the zero-order Bessel function 
of the first kind could also be  expanded so that the conduction and valence 
bands $(\gamma= \pm 1)$ energy dispersions are further approximated as 

\begin{equation} 
 \varepsilon_d^{\,(L)} (\phi \, \vert \, \lambda_0, {\bf k}) \backsimeq \pm \hbar v_F \,k \,\, 
 \left\{ 1  - \frac{\lambda_0^2}{8} \, \left[ 5 + 3 \cos (4 \phi) \right]
\, \sin^2 \theta_{\bf k} 
\right\} \, .
\end{equation}
Even for an infinitesimal coupling constant $\lambda_0$, the anisotropy and, 
therefore, the difference between Fermi velocities in the $k_x$- and 
$k_y$-directions is the largest for graphene and the smallest in the dice lattice
case. 

\begin{figure}
\centering
\includegraphics[width=0.49\textwidth]{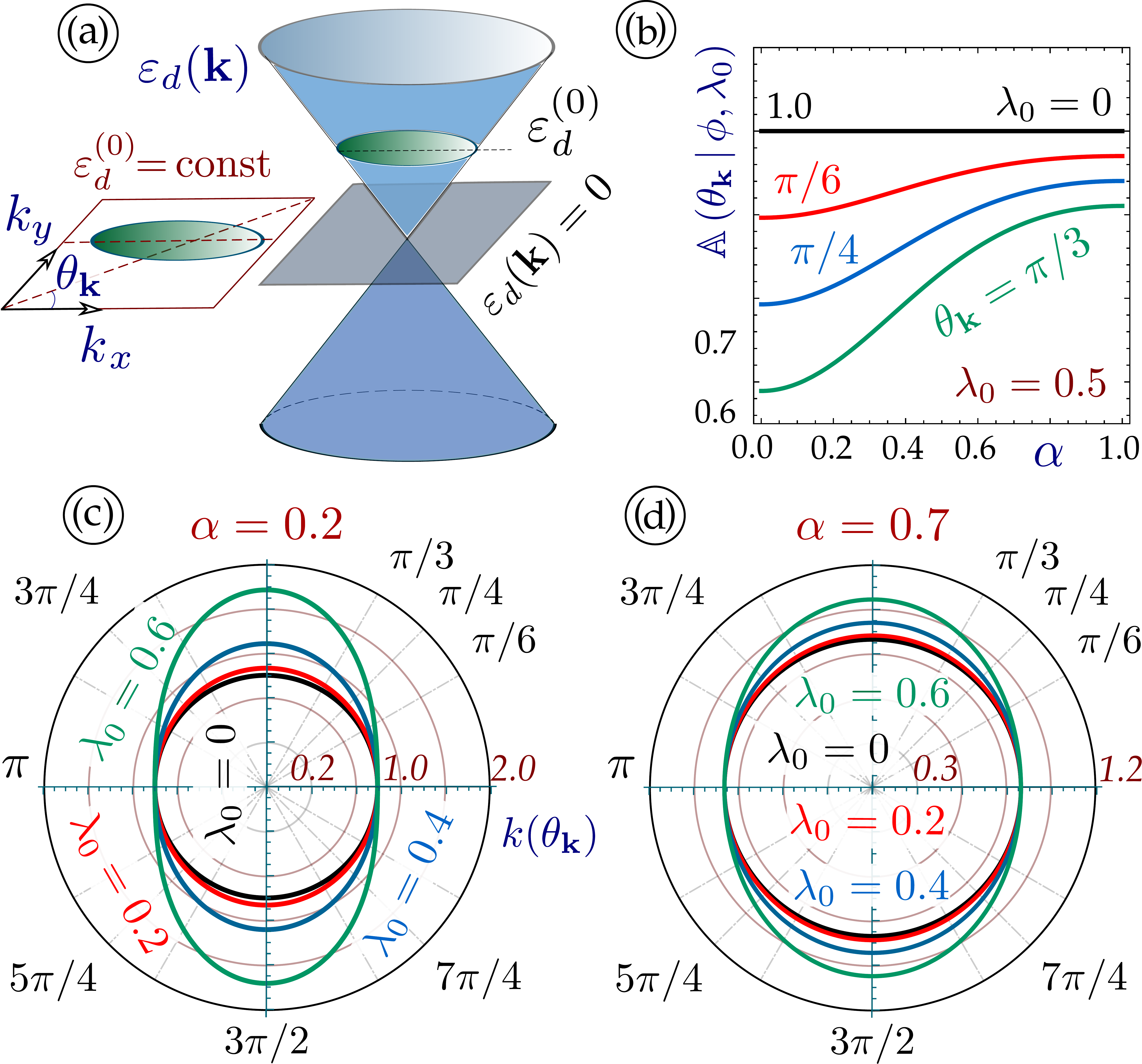}
\caption{(Color online) Angular dependence of dressed-state energy 
dispersions $\varepsilon_d^{\,(L)} (k, 
\theta_{\bf k})$, $\theta_{\bf k} = \tan^{-1}\,\left(k_y / k_x \right)$  
for the case of linearly-polarized irradiation applied to an $\alpha$-T$_3$ 
lattice, as it is schematically shown in panel $(a)$. Plot $(b)$ represents 
how the angular factor $\mbb{A}(\theta_{\bf k} \, \vert \, \phi, \lambda_0)$, given by
Eq.~\eqref{TA}, depends on lattice parameter $\alpha$ at various 
angles $\theta_{\bf k}$ ($\theta_{\bf k}
= \pi/6$ corresponds to a red line, $\pi/4$ - blue curve and 
$\theta_{\bf k} = \pi/3$ is depicted by the green 
line). The two lower polar plots $(c)$ and $(d)$ show the angular 
dependence of the obtained anisotropic dispersions \eqref{linD} for 
chosen energy $\varepsilon_0 = 0.5 \,E_F$. In each plot, the  
curves are related to the different coupling parameters $\lambda$, 
as labeled. Here,  $\lambda = 0$ is used for the black lines, 
$\lambda = 0.2$ for the red curves, $\lambda = 0.2$  for blue 
ones and $\lambda = 0.6$  for the green lines.}
\label{FIG:5}
\end{figure}
 
\par 
\medskip 
\par 

First and most important, we see that the flat band $\varepsilon({\bf k}) = 0 $ 
is not affected and remains dispersionless under linearly-polarized irradiation for 
all wave vectors. The valence and conduction bands become anisotropic,
 so that the standard right-circular Dirac cone is transformed into an elliptic 
cone with its major axis located along the direction of the light 
polarization. We also notice that the complete electron-hole symmetry 
of the upper and lower cones is preserved and there is no energy gap 
between the valence and conduction bands. These features are  completely
similar to the corresponding results for graphene.\cite{kisrep}    
However, an important novelty comes from the $\phi$-dependence of 
the obtained dispersions. As we demonstrate in Fig.~\ref{FIG:4},  
the strength  of induced anisotropy is related to
the specific value of chosen $\phi$. 

\par 
\medskip
\par 

Even though the procedure of deriving the corresponding wave functions, 
explained in Appendix \ref{apb}, is straightforward, the  calculations
are lengthy and  tedious even for the simplest case of a dice lattice. The 
two wave functions, corresponding to the valence and conduction band 
energies with $\gamma = \pm 1$ are as follows

\begin{eqnarray}
\label{Twlfd}
 && \Psi_d^{\gamma = \pm 1} (\lambda_0, {\bf k}) = \tet{exp}[ \mp i v_F k t \, f_\theta] \, 
 \frac{f_\theta + \cos\theta_{\bf k}}{4 \,f_\theta} \, \times \,
  \left[
 %% T1
 \tet{e}^{\pm i z_\lambda(t)} \,
 \left\{  
 \begin{array}{c}
  \tau \\
  \pm \sqrt{2} \\
  \tau
 \end{array}
 \right\} - \right. \\ 
 \nonumber 
&& \left.
% T2
  - \, 2 i \, \frac{\mc{J}_0(\lambda_0) \, \sin \theta_{\bf k}}{f_\theta + \cos \theta_{\bf k}} \,
\left\{  
 \begin{array}{c}
  1 \\
  0 \\
  - 1
 \end{array}
 \right\} -
 % T3
 \left[ 
 \frac{\mc{J}_0(\lambda_0) \, \sin \theta_{\bf k} }{f_\theta + \cos \theta_{\bf k}}
 \right]^2
 \left\{  
 \begin{array}{c}
  \tau \\
  \mp \sqrt{2} \\
  \tau
 \end{array}
 \right\} \, \tet{e}^{ \mp i z_\lambda (t)}
 \right] \, ,
\end{eqnarray}
where 

\begin{equation}
\label{Tdrf}
 f_\theta = \sqrt{\,\mbb{A}(\theta_{\bf k} \, \vert \, \phi = \pi/4, \lambda_0) } =
 \left\{ 
 \cos^2 \theta + \left[
 \sin \theta \,
 \mc{J}_0 (\lambda_0) \,
 \right 
 ]^2 \,
 \right\}^{1/2}
\end{equation}
and $z_\lambda(t) = \lambda_0 \, \sin (\omega t)$. Once its components are 
directly calculated, the wave function \eqref{Twlfd} takes the following form 

\begin{eqnarray}
\label{new1}
 && \Psi_d^{\gamma = \pm 1} (\lambda_0, {\bf k}) = \frac{1}{2} \, 
 \left\{
 \begin{array}{c}
  \tau \, \tet{e}^{-i \Phi_\theta (\lambda_0)} \\
  \gamma \, \sqrt{2} \\
  \tau \, \tet{e}^{i \Phi_\theta (\lambda_0)}
 \end{array}
 \right\} \, , \\
 \nonumber 
 && \Phi_\theta (\lambda_0) = 2 \arctan \left[ \tau \,
 \frac{\mc{J}_0(\lambda_0)\,\sin \theta_{\bf k}}{f_\theta + \cos \theta_{\bf k}} 
 \right] \backsimeq \tau \, \left[ \theta_{\bf k} - \frac{\lambda_0^2}{8} \, 
 \sin^2 (2  \theta_{\bf k}) + ... \right] \, .
\end{eqnarray}
The components of this wave function are apparently equal to each other 
up to a phase difference similar to the non-interacting dice lattice wave functions, 
which corresponds to Eq.~\eqref{Eig1} and \eqref{Eig1} for $\phi=\pi/4$. The 
phase difference $\Phi_\theta (\lambda_0)$  for a dice lattice is no longer equal 
to $\theta_{\bf k}$ and depends on the imposed irradiation intensity. 

\par 
\medskip 
\par 

The remaining wave function for the flat band with $\gamma = 0$ is given by 

\begin{equation}
\label{Twfld0}
 \Psi_d^{0} (\lambda_0, {\bf k}) = \frac{1}{\sqrt{2} \,f_\theta} \, \times \, 
\left[ 
 - \frac{i \tau}{2} \, \sin \theta_{\bf k} \, \mc{J}_0(\lambda_0) \,
  \sum\limits_{\alpha = \pm 1}
 \tet{e}^{i \alpha \, z_\lambda(t)} \,
 \left\{  
 \begin{array}{c}
  1 \\
  \alpha \sqrt{2} \, \tau  \\
  1
 \end{array}
\right\}  + \cos \theta_{\bf k} \, 
\left\{  
 \begin{array}{c}
  1 \\
  0 \\
  - 1
 \end{array}
 \right\}
\, \right] \, .
\end{equation}
The structure of the obtained wave function in Eq.\ \eqref{Twfld0} is 
such that it again consists of two non-zero components of equal amplitude
 and phase difference $\Phi_\theta (\lambda_0)$. Consequently, we can 
rewrite the wave function \eqref{Twfld0} as 

\begin{eqnarray}
\label{new2}
&& \Psi_d^{0} (\lambda_0, {\bf k}) = \frac{1}{\sqrt{2}} \, 
\left\{ 
\begin{array}{c}
 \tet{e}^{- i \Phi_\theta (\lambda_0)} \\
 0 \\
 - \tet{e}^{i \Phi_\theta (\lambda_0)}
\end{array}
\right\} \, , \\
\nonumber 
&& \Phi_\theta (\lambda_0) = \arctan \left[ 
\tau \, \mc{J}_0(\lambda_0) \, \tan \theta_{\bf k}
\right] \backsimeq \tau \left[
\theta_{\bf k} - \frac{\lambda^2}{2} \, \sin (2 \theta_{\bf k}) + ...
\right] \, .
\end{eqnarray}
It is interesting to compare our results with the corresponding wave function 
for graphene obtained in Ref.~[\onlinecite{kisrep}]. The Dirac electron in 
graphene, interacting with a linearly-polarized off-resonant dressing field has 
the following dispersions 

\begin{eqnarray}
 && \varepsilon_d^{\gamma = \pm 1} = \gamma \hbar v_F k \times f_\theta \, , 
 \\ 
 \nonumber 
 && f_\theta = \left\{ 
 \cos^2 \theta_{\bf k} + \left [
 \mc{J}_0 (2\lambda_0) \, \sin \theta_{\bf k} 
 \right]^2 \,
 \right\}^{1/2} \, . 
\end{eqnarray}
The anisotropy factor $f_\theta $ for graphene is equivalent to our 
expression \eqref{rsf} for $\phi$ or $\alpha = 0$. This is an opposite limit 
for the angular dependence in Eq.~\eqref{rsf} from the dice lattice with 
$\phi = \pi/4$, given in Eq.~\eqref{drf}.

\par 
\medskip 
\par 

The corresponding wave function at $t=0$ could be rewritten as 

\begin{eqnarray}
\label{wfgl}
&& \Psi_d^{\gamma = \pm 1}({\bf k}) = \frac{1}{\sqrt{2}} \, \left\{ 
\begin{array}{c}
 1 \\
 \gamma \, \tet{exp} \left[ i \, \Phi_\theta(\lambda_0) \right]
\end{array}
\right\} \, , \\
\nonumber 
&& \Phi_\theta(\lambda_0) = 2 \arctan \left[ \frac{
\sin \theta_{\bf k} \, \mc{J}_0(\lambda_0)
}{
\cos \theta_{\bf k} + f_\theta
} \right] \backsimeq  \theta_{\bf k} - \frac{\lambda_0^2}{2} \, \sin (2 \theta_{\bf k}) + \, ... \,
 \, .
\end{eqnarray}
As it is expected for a distorted anisotropic Dirac cone, its wave function 
\eqref{wfgl} has equal components, while their phase difference 
$\Phi_\theta(\lambda_0)$ depends on the interaction coefficient $\lambda_0$.

\par 
\medskip 
\par

For the general case of an $\alpha$-T$_3$ lattice, the wave function is 
obtained as 

\begin{eqnarray}
\label{Twfl}
 && \Psi_d^{\gamma = \pm 1} (\lambda_0, {\bf k}) =  
\frac{1}{
\left[
 \mc{N}_{23}^{\,(\phi)}
 \right]^{1/2}
}
\, \times \,
 \tet{exp}\left[ \mp i v_F k t \, f_\theta^{\,(\phi)}
 \right] \, \times
 \\ 
 \nonumber 
&& \times \, \left[
 %% T1
 \frac{r_1^{\,(\phi)}}{\sqrt{2}} \,
 \left\{  
 \begin{array}{c}
  \tau \, \cos \phi\\
  \pm 1\\
  \tau \, \sin \phi
 \end{array}
 \right\} \,  \tet{e}^{\pm i z_\lambda(t)} + 
% T2
 r_2^{\,(\phi)} \,
\left\{  
 \begin{array}{c}
  \sin \phi \\
  0 \\
  - \cos \phi
 \end{array}
 \right\} +
 % T3
\frac{r_3^{\,(\phi)}}{\sqrt{2}} \,
 \left\{  
 \begin{array}{c}
  \tau \cos \phi\\
  \mp 1 \\
  \tau \sin \phi
 \end{array}
 \right\} \, \tet{e}^{ \mp i z_\lambda (t)}
 \right] \, ,
\end{eqnarray}
where the coefficients $r_{1,2}^{\,(\phi)}$ and the normalization function 
are defined in the Appendix \ref{apb}.

\par 
\medskip 
\par 

For the flat band with $\gamma = 0$, the wave function takes the form

\begin{equation}
\label{Twfl0}
 \Psi_d^{0} (\lambda_0, {\bf k}) = \frac{1}{ 
 \sqrt{
 2 \, \mc{N}_{\theta}^{\,(\phi)} 
 }
 }
 \, \times \, 
\left[ 
 r_1^{\,(\phi)} \, \frac{\tau}{\sqrt{2}} \, \sum\limits_{\alpha = \pm 1}
 \tet{e}^{i \alpha \, z_\lambda(t)} \,
 \left\{  
 \begin{array}{c}
  \cos \phi \\
  \alpha \, \tau  \\
  \sin \phi
 \end{array}
\right\}  + r_2^{\,(\phi)} \, 
\left\{  
 \begin{array}{c}
  \sin \phi \\
  0 \\
  - \cos \phi
 \end{array}
 \right\}
\, \right] \, , 
\end{equation}
where 

\begin{eqnarray}
\label{Trsf}
 && r_1^{\,(\phi)}(\lambda_0, {\bf k}) = - i \sin (2 \phi) \, \sin \theta_{\bf k} \, \mc{J}_0 (\lambda_0) \, , \\
 \nonumber 
 && r_2^{\,(\phi)}(\lambda_0, {\bf k}) =  \sqrt{2} \, \left[ 
 \cos  \theta_{\bf k} + i \tau \, \cos (2 \phi) \, \sin \theta_{\bf k} \, \mc{J}_0 (2\lambda_0)
 \right] \, .
 \nonumber
\end{eqnarray}
The components of the wave function \eqref{Twfl0} are not equal to each 
other since this condition does not hold true even in the absence of irradiation, 
while all the previously obtained wave functions have  components which  
differ only by a phase factor. This becomes explicit once the components of 
this wave function are evaluated 

\begin{eqnarray}
 && \Psi_d^{0} (\lambda_0, {\bf k}) = \frac{1}{\sqrt{\mc{N}_{\theta}^{\,(\phi)}}} \,  
\left\{ 
\begin{array}{c}
 \sin \phi \, \left[
 \cos \theta_{\bf k} - i \tau \, \sin \theta_{\bf k} \, \mbb{X}^{\,(\phi)}_\theta (\lambda_0) \,
 \right] \\
 0 \\
 - \cos \phi \, \left[
 \cos \theta_{\bf k} + i \tau \, \sin \theta_{\bf k} \, 
 \left( 2 \, \mc{J}_0 (\lambda_0) -
 \mbb{X}^{\,(\phi)}_\theta (\lambda_0) 
 \, \right) \,
 \right]
\end{array}
\right\} \, ,
\end{eqnarray}
where, in our situation with $\lambda_0 \ll 1$, we can write 

\begin{eqnarray}
&& \mbb{X}^{\,(\phi)}_\theta (\lambda_0) \backsimeq 1 - \frac{\lambda_0^2}{4} \,
\left[ 4 - 3 \cos (2 \phi) \right] \, , \\
\nonumber 
&&  2 \, \mc{J}_0 (\lambda_0) -
 \mbb{X}^{\,(\phi)}_\theta (\lambda_0) \backsimeq  1 - \frac{\lambda_0^2}{4} \,
\left[4 + 3 \cos (2 \phi) \right] \, , \\
\nonumber
&& \mc{N}_{\theta}^{\,(\phi)} \backsimeq 1 - \frac{\lambda_0^2}{8} \left\{
5 - \left[
5 \cos (2 \, \theta_{\bf k}) - 6 \cos (4 \phi) \, \sin^2 \theta_{\bf k}
\right] \,
\right\} \, .
\end{eqnarray}
 The components of this wave function are not equal to each other, in contrast 
to all previously considered cases involving linearly-polarized field. The way they 
are modified in the presence  of electron-photon interaction is not correlated 
with their initial values.  Surprisingly, the normalization factors for all the obtained 
wave functions \eqref{Twlfd}-\eqref{Twfld0} and
\eqref{Twfl}-\eqref{Twfl0} do not depend on time.

\section{Berry phase modification due to the dressing field}
\label{s3}

As an evident and natural application for our novel photon-dressed electronic 
states obtained in Sec.~\ref{s2}, we consider how the Berry phase of an $\alpha$-T$_3$ or 
dice lattice is affected by the presence of an off-resonance 
dressing field with different polarizations. Specifically, we are interested in
examining the way in which such quantum phases depend on $\alpha$ in 
the presence of finite electron-light coupling $\lambda_0 > 0$.

\par 
\medskip 
\par 

The Berry phase is defined as a geometrical phase difference, which a 
purely quantum system receives over a complete cycle of adiabatic, or 
isoenergetic evolution.\,\cite{A127, A129, A1Thesis} All  physically meaningful
parameters, except for a quantum phase, are expected to return to their 
initial values over such a loop-like transformation. The Berry phase is logically connected to the quantal 
phase of \textit{Aharonov and Bohm effect} for a 
charged particle moving along a closed path, which partially includes 
electrostatic or magnetic fields.  Such topological phases strongly affect 
transport properties and lead to a finite conductivity at the band crossing
even if the density of propagating waves at this point vanishes.\,\cite{A131}

\par
\medskip
\par

As the first step, we are going to carry out a detailed investigation of the 
time-independent eigenstates, corresponding to the distorted Dirac cone 
due to an electron interacting with a dressing field. This could lead to either 
the opening of a band gap in the case of an elliptically- or circularly-polarized 
field, or an anisotropy of the dispersion relations due to a field with linear 
polarization. While in the former case the energy dispersions \eqref{SpecRes} 
were obtained using an effective perturbation Hamiltonian \eqref{Tmexp} 
without taking its actual time dependence into account, the wave functions 
for the latter case involving linearly-polarized light were obtained with complete
$t-$dependence, and therefore, need to be further clarified.  

\par 
\medskip

This could be viewed as a simplified model with an additional $\hat{\Sigma}_z$ 
term in graphene under circularly-polarized light\,\cite{ourJAP2017}, which leads 
to considerable explanation of the considered phenomena, yet bears nearly 
all the crucial properties of irradiated graphene.\,\cite{kisrep} We will suppress the 
time dependence in Eqs. \eqref{Twfl} and \eqref{Twfl0} and use their expressions 
at $t=0$ for all further considerations. This will allow us to address a wider class 
of field-induced electronic states, not necessarily equivalent to dressed states 
under conditions discussed above.

\par 
\medskip
\par 
In general, the time dependence of the obtained eigenstates, such as Eqs.
\eqref{Twfl} and \eqref{Twfl0},is expressed in exponential dependence of 
some of their components $\tet{exp}[\pm i \, z_\lambda(t)] =
\tet{exp}[\pm i \, \lambda_0 \sin (\omega t)]$. In our consideration for an 
off-resonant field with $\lambda_0 \ll 1$, this dependence would lead to 
noticeable modification of the wave function components.  Surprisingly, the normalization factors $\mc{N}_\theta^{\,(\phi)}$ for all types of 
$\alpha$-T$_3$ materials under linearly-polarized irradiation, including the 
dice lattice limit, do not directly depend on time. Another occasion of direct time dependence is the initial phase factor $ \tet{exp}\left[ \mp i v_F k t \, 
f_\theta^{\,(\phi)} \right]$ in Eqs.~\eqref{Twfl} and \eqref{Twlfd}, which does 
not affect most of its properties but leads to a linear increase of its Berry phase 
over time as $\backsim v_F t \, f_\theta^{\,(\phi)}$.

\par 
\medskip
\par 

The described situation of the time dependence is strikingly similar to the \textit{electrostatic Aharonov-Bohm interference} effect.\cite{A1AB} The 
wave function of a conventional Schr{\"o}dinger particle of energy 
$\mbb{E}$ acquired a phase factor $\tet{exp}[i \phi]$, $\phi = - \mbb{E} t/\hbar$. 
If such a particle is confined in the region with a constant electrostatic potential 
$V_0 = $ constant, and zero electrostatic field, this potential affects the phase 
of the eigenstate as $\phi(t)-\phi(t=0) =-e V_0 t /\hbar$. This additional phase 
affects the actual properties of the particle, as well as the  outcome of a 
double-slit interference experiment, in spite of the fact that the actual 
electrostatic field is not present. 

\par
\medskip
\par
We can write the Berry phase as \,\cite{A129, A1Thesis} 

\begin{equation}
\label{Berry}
\phi_B (\tau)=-i \,\oint_\mbb{C} d {\bf k} \cdot \left\{
\left[ 
 \Psi_{\tau,\,\phi}^{\gamma}({\bf k}) \right]^\dag \vec{\nabla}_{{\bf k}}
 \Psi_{\tau,\,\phi}^{\gamma}({\bf k})
 \right\} \, ,
\end{equation}
where $\mbb{C}$ is a closed path within the  2D plane. For the case of 
non-irradiated wave functions of an $\alpha$-T$_3$ lattice presented in Eqs.\ \eqref{Eig1} and \eqref{Eig2}, the result is immediately obtained as 
$\phi_{\gamma = \pm 1,\tau}^B=\tau \pi \, \cos(2\phi)$ for the conduction 
and valence bands, and $\phi_{0 ,\tau}^B=2\tau \pi \, \cos(2\phi)$ for the flat 
band at the $K$ and $K^\prime$ valleys. These results \textit{do not depend} 
on the choice of a closed curve $\mbb{C}$, which in general cannot be true
since the wave function components are $k-$dependent. We also remark that the 
Berry phase is gauge invariant and its value is  unique up to $2\pi \times $ integer.

\begin{figure}
\centering
\includegraphics[width=0.49\textwidth]{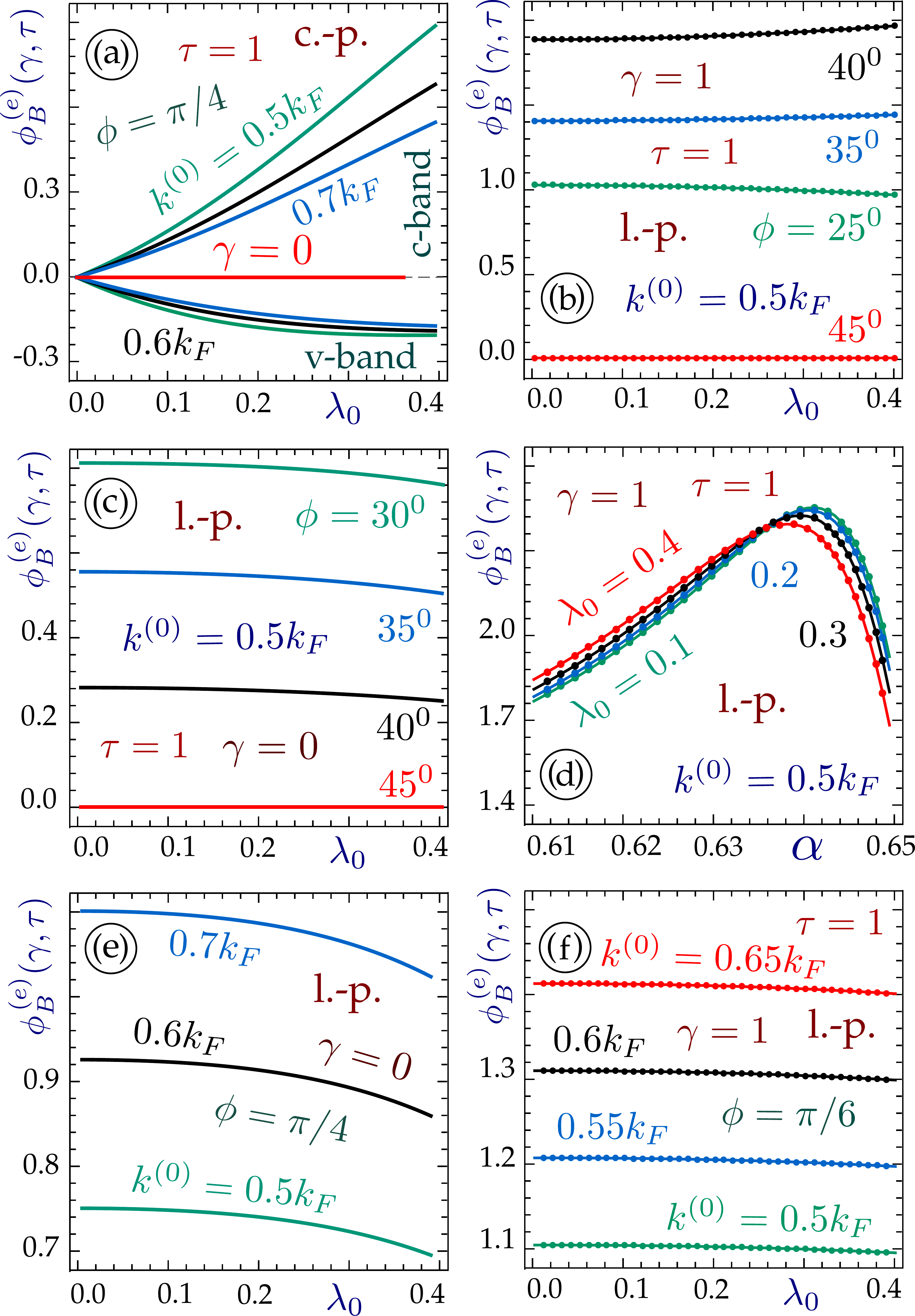}
\caption{(Color online) Berry phase for irradiated $\alpha$-T$_3$ 
materials. Panel $(a)$ corresponds to the circular polarization of the
 incident light, while all the others $(b)$-$(f)$  are related to the linear type. 
Panel $(a)$ shows the results for all bands, $\phi^{\,(e)}_B > 0$ for the 
conduction band,  $\phi^{\,(e)}_B < 0$ for the valence band, and the values 
of Berry phase are not symmetric. However, it is always zero for the flat band.  
The remaining  plots $(c)$ and $(e)$ on the left are related to the flat band of an 
$\alpha$-T$_3$ lattice, and all the right panels $(b)$, $(d)$ and $(f)$
 to the conduction band with $\gamma = 1$.  All plots except $(d)$ show 
the Berry phase dependence on the coupling constant $\lambda_0 < 1$, but 
$(d)$  shows a dependence on the parameter $\alpha = \tan \phi$ of an 
$\alpha$-T$_3$ material.  In panels $(a)$ and $(e)$, each black curve 
corresponds to $0.5\,k_0$, the red curve to  $0.6\,k_0$ and the blue curve
to $0.7\,k_0$. Blue, black and red curves are related to $k^{(0)}=0.55\,k_0$, $0.6\,k_0$ and $0.55\,k_0$ in plot $(f)$.  Each line in panels $(b)$ and $(c)$ 
corresponds to a specific value of $\phi$ as labeled. In the case  of a dice 
lattice with $\phi = \pi/4$, $\phi^{\,(e)}_B = 0$ disregarding  the electron-light coupling. }
\label{FIG:6}
\end{figure}

The details on how to evaluate linear integral in Eq. \ \eqref{Berry} in 
polar coordinates in  $k-$space for various types of incoming light polarization 
are provided in Appendix \ref{apc}. In our first case of circularly-polarized light, 
components  \eqref{Twf1} of the wave function \eqref{AAwf01} are only 
$k$-dependent except for $\tet{e}^{\pm i \theta_{\bf k}}$ factors, which are 
similar to those in the non-irradiated eigenstates \eqref{Eig1} and \eqref{Eig2}, 
and the path of such isoenergetic linear integration is a circle of radius 
$k^{\,(0)}$. As a result, the Berry phase in the case of a dice lattice
irradiated by a circularly-polarized dressing field is given by the following closed-form analytical expression 

\begin{equation}
\label{TB1}
 \phi^{\,(C)}_B (\tau = 1) = \frac{
 \left\{ \mc{C}^{(1)} \left[ k^{\,(0)} \right] 
 \right\}^2 - 1
 }{
 \mc{N}^{\,(e)} \left[ k^{\,(0)} \right]
 } \,
 \int\limits_{0}^{2 \pi} d \theta_{\bf k} \backsimeq \pi \, \frac{c_0}{\hbar v_F k^{\,(0)}}  \, \lambda_0 \, 
 \left[
 \gamma + \frac{c_0}{\hbar v_F k^{\,(0)}}  \, \lambda_0
 \right] \, .
\end{equation}
The obtained Berry phases are not symmetric and are  opposite for the 
electron and hole states except for the first-order term in the $\lambda_0$ 
expansion, as we also see from Fig.~\ref{FIG:6}$(a)$, even though this 
expression is exactly symmetric over the valley index $\tau = \pm 1$, i.e., $\phi^{\,(C)}_B (\tau = - 1) =
- \phi^{\,(C)}_B (\tau = 1)$, which is in complete analogy with the dice lattice in the absence of  irradiation.\,\cite{A1Thesis} In summary, even for a dice lattice that does not have a finite Berry phase ($\phi^{\,(0)}_B (\tau) \backsim \cos (2 \phi) = 0$), it  can be generated by a gap-opening elliptically- or circularly-polarized dressing field due to the variation of wave function components.  While the dispersions are symmetric 
for a dice lattice, the wave function components, and therefore, the
Berry phases do not share this property.
The corresponding phase for the flat band is immediately obtained as zero, 
irrespective of the valley index or intensity of the incoming radiation since only the middle component in the  eigenstate $\eqref{Twf2}$ is affected
by the electron-light interaction, gives no contribution to Eq.~\eqref{Berry}. 

\par 
\medskip 
\par 

Finally, we have calculated numerically the Berry phase for the 
$\alpha$-T$_3$ lattice with an arbitrary $\phi$ in the presence
of linearly-polarized irradiation. Our results are presented in Fig.~\ref{FIG:6}. 
Panels $(c)$ and $(e)$ represent the results for the flat band, and the 
three right plots $(b)$, $(d)$ and $(f)$  for the  conduction band with 
$\gamma = 1$. In both cases, the phase is zero for a dice lattice, disregarding  
the light intensity or the parameters of isoenergetic elliptic integral path so 
that the results in panel $(d)$ are equal to zero in the limit $\alpha \to 1$, which confirms our previous analytical results obtained previously in Appendix \ref{apc} 
for a dice lattice under dressing field with linear polarization. The results for the 
flat band demonstrate a stronger dependence on the coupling constant, as well 
as on the parameters for an elliptic path of the linear integration. As we see from 
panel $(d)$ of Fig.~\ref{FIG:6}, the dependence of the Berry phase on $\lambda_0$ 
is not monotonic throughout the  considered values of $\alpha$, namely, the 
order in which all the curves corresponding to different $\lambda_0$
are laid out changed to exactly opposite ones when each line begins, 
passes its maximum value and starts  decreasing as a function of 
$\alpha$. Such non-monotonic dependence has not been observed for
 the case of circularly-polarized irradiation. 

\section{Concluding remarks}
\label{s4}

In this paper, we have executed a thorough investigation into
electron-photon dressed states in $\alpha$-$T_3$  lattices
for all possible polarizations (elliptical, circular and linear) of the
 impinging radiation. We have derived closed-form expressions
and analytic approximations for the quasiparticle energy dispersions 
for all types of such optical states.  

\par 
\medskip
\par 

We have demonstrated that the  phase $\phi$ or the hopping parameter 
$\alpha$ plays a crucial role and affects the low-energy band structure for 
each type of polarization of the incident light. The obtained dressed states 
demonstrate both similarity and strong distinction compared to those earlier 
obtained in graphene or buckled honeycomb lattices. As an example, 
elliptically-polarized irradiation is connected with opening a band gap in the 
energy dispersions of $\alpha$-T$_3$, symmetry breaking between the valence 
and conduction bands. The parameter $\alpha$ has also  been shown to 
strongly affect the radiation-induced anisotropy and the angular dependence of 
the dressed quasiparticle dispersions for the case of a linearly-polarized external 
field. Generally speaking, we find that the effect of electron-light coupling is strongest 
for $\alpha \to 0$ (graphene), and is the weakest for a dice lattice with $\alpha = 1$ 
for both types of incoming light polarization.

\par 
\medskip
\par 

We have found that for an elliptically-polarized field applied to a material 
with $\alpha \neq 1$, its low-energy band structure, including the gaps, directly
depends on the $\tau = \pm 1$ valley index. This gives rise to valleytronics applications,
which now could also  be developed based on the irradiated $\alpha$-T$_3$ lattice and enables
electrically controllable valley filtering as being crucial for such applications and 
technology.

\par 
\medskip
\par

In addition to calculating the energy dispersion relations, we have 
obtained analytically the corresponding wave function for electrons
dressed by an external field of various polarization. For elliptically-polarized 
light, the components of the obtained eigenstates are not equal to each 
other and do not just vary by a phase factor, as we have in the absence
of irradiation \eqref{Eig1} and \eqref{Eig2}. These components directly 
depend on the wavevector $k$ and on  the band index $\gamma$, so that 
the modification of the eigenstates in the valence, conduction and flat bands 
is not the same. This is the first obtained wave function for a dice lattice 
with a finite energy band gap. 

\par 
\medskip
\par

Unlike previously discussed circularly-polarized irradiation, the 
eigenvalue equation for the linear polarization of the incident field 
allows for an exact solution for ${\bf k}$, so that we can claim with 
complete precision that in this case there is no energy gap between 
the conduction and valence bands, the flat band permanently stays 
dispersionless at the zero-energy level and complete symmetry 
between the bands is conserved. The effect of linearly-polarized 
dressing leads to the anisotropy of the Dirac cone dispersions similar
to graphene. However, now we are able to tune the anisotropy and angular 
dependence of the energy band structure by adjusting the parameter 
$\alpha$. Apart from the band symmetry breaking in the irradiated 
$\alpha$-T$_3$ model reported in Ref.~[\onlinecite{hu}], we reveal a 
number of other crucial symmetries of the obtained dressed states, which 
could be broken or persist depending on the type and intensity of the 
applied field and the value of $\alpha$.

\par 
\medskip
\par

The corresponding wave functions are drastically different for the dice lattice 
with $\alpha = 1$ and all other possible $\alpha$-T$_3$ materials. In the former 
case, the components of such a dressed state are equal and differ only by a phase 
factor, which similar to anisotropic Dirac fermions in few-layer black 
phosphorus \cite{liklein} are expected to reveal non-head-on asymmetric 
Klein paradox. In contrast, the components of such a wave function
for  arbitrary $0 < \alpha < 1$ differ from each other, which brings in 
considerable modification of their tunneling and transport properties. For a number 
of evaluations, we consider time-independent wave functions corresponding 
to $t = 0$ in order to address a wider class of phenomena pertaining to 
light-induced distortions of Dirac cone dispersions for $\alpha$-T$_3$ lattices.

\par
\medskip 
\par 

We also investigated Berry phases of the obtained dressed electron eigenstates 
in connection with their unusual composition and symmetric properties. The 
Berry phase is a specific quantum characteristic of an electronic state,
which is particularly sensitive to a particle's environment and adiabatic 
change of external fields or their potentials. Such phases could be sometimes 
acquired by a system under consideration even if all the other important
parameters and quantum numbers remain unaltered.

\par
\medskip 
\par 

Berry phases are directly related to the wave function it is attributed to,
its components, and the phase difference between them. For instance, 
the phases corresponding to the valence and conduction bands for  
  a gap opening elliptically- or circularly-polarized irradiation even for
 the simplest dice lattice are not symmetric of opposite to each other, 
unlike their energy dispersions. We uncovered these relations in all their 
complexity and demonstrated that the phase of the flat band is always 
zero. The same is true for the phases of a dice lattice eigenstates for 
all bands ($\gamma = \pm 1$ and $\gamma = 0$) in the presence 
of an external field with linear polarization. For all other values of 
$\alpha \neq \pi/4$, we observed their moderate dependence on 
electron-light coupling $\lambda_0$, parameter $\alpha$ and on the 
closed integration path, as it is expected to be according to 
the Aharonov-Bohm effect. The modification of the Berry phases 
pertaining to specific coupled electron-light states affect some of 
their important properties and could considerably modify the results 
of double-slit interference observations.

\par 
\medskip
\par 

By studying the band structure of such optical electron states, we 
developed a useful methodology for laser-induced engineering of 
energy bands and tuning their most significant characteristics of 
$\alpha$-T$_3$ innovative materials. Our results are expected to 
have a profound effect on the fabrication of modern optical and 
electronic devices and photonic crystals.  

\acknowledgments
D.H. would like to acknowledge the support from the Air Force Office 
of Scientific Research (AFOSR). D.H is also supported by the DoD Lab-University 
Collaborative Initiative (LUCI) program.
G.G. would like to acknowledge the support from the Air Force
Research Laboratory (AFRL) through Grant \#12530960.

\appendix

\section{Electron-field dressed states for elliptically- and circularly- polarizaitons
applied to a dice lattice ($\phi = \pi/4$) - derivation of Eqs. \eqref{SpecRes}, 
\eqref{Twf1} and \eqref{Twf2}}.

\label{apa}

For a dice lattice, the time-independent perturbation operator \eqref{TP} is expressed 

\begin{equation}
 \hat{\mbb{P}}_{\, \tau} = - \frac{\tau \,c_0}{2\sqrt{2}} \, \sum\limits_{\alpha = \pm}
 (1 - \alpha \tau \, \beta) \, \hat{\Sigma}_{\alpha} \, 
\end{equation}

in terms of 
 
\begin{equation}
\Sigma^{\,(1)}_+ = 
\left( 
\begin{array}{cc} 
    \begin{array}{c} 
    0 \\ 0 
    \end{array}   
    & \mbb{I}_{2 \times 2}  \\
    0 & \begin{array}{cc}
    0 & 0 
    \end{array}
\end{array}
\right) = 
\left(
\begin{array}{ccc}
 0 & 1 & 0 \\
 0 & 0 & 1 \\
 0 & 0 & 0
\end{array}
\right) 
\hspace{0.3in} 
\text{and}
\hspace{0.3in} 
\Sigma^{\,(1)}_- = 
\left[ 
\begin{array}{cc}     
        \begin{array}{cc}
    0 & 0 
    \end{array}
    & 0  \\    
    \mbb{I}_{2 \times 2} &     
    \begin{array}{c} 
    0 \\ 0 
    \end{array}   
    \end{array}
\right] = 
\left[
\begin{array}{ccc}
 0 & 0 & 0 \\
 1 & 0 & 0 \\
 0 & 1 & 0
\end{array}
\right] \, .
\end{equation}
The operators $\hat{\Sigma}^{\,(1)}_{\pm}$ are built from the
 spin-$1$ matrices 

\begin{equation}
 \hat{\Sigma}_x^{\,(1)} = \frac{1}{\sqrt{2}} \, \left(
 \begin{array}{ccc}
  0 & 1 & 0 \\
  1 & 0 & 1 \\
  0 & 1 & 0
 \end{array}
 \right)
\hspace{0.3in} 
\text{and}
\hspace{0.3in} 
 \hat{\Sigma}_y^{\,(1)} = \frac{1}{\sqrt{2}} \,
 \left(
 \begin{array}{ccc}
  0 & -i & 0 \\
  i & 0 & -i \\
  0 & i & 0
 \end{array}
 \right) \, ,
\end{equation}

where $\hat{\Sigma}^{\,(1)}_{\pm} = \hat{\Sigma}^{\,(1)}_{x} \pm i \hat{\Sigma}^{\,(1)}_{y}$, 
similarly to the  case of  $2\times 2$ Pauli matrices for spin $1/2$ used for graphene.
Similarly, the energy gap exists if a $\hat{\Sigma}^{\,(1)}_{z}$ matrix is present in 
the Hamiltonian. In our case, this matrix is expressed as 

\begin{equation}
\hat{\Sigma}_z^{\,(1)} = \frac{1}{\sqrt{2}} \,
 \left(
 \begin{array}{ccc}
  1 & 0 & 0 \\
  0 & 0 & 0 \\
  0 & 0 & -1
 \end{array}
 \right) \, . 
\end{equation}

Finally, the effective pertubation Hamiltonian \eqref{TeffH} is modified in the following way 

\begin{equation}
\label{AHam}
\hat{\mc{H}}_{\text{eff}}^{\,(E)} ({\bf k},t) = 
\frac{\hbar v_F}{\sqrt{2}}\sum\limits_{\alpha = \pm} \left(\tau k_x - \alpha i  k_y  \right) \, \Sigma_\alpha^{\,(1)}  -
\frac{\tau \beta}{2} \, \lambda_0 \, \hat{\Sigma}_z^{\,(1)} + \hbar v_F \frac{\tau}{4\sqrt{2}} \, \lambda_0^2 \,
\sum\limits_{\alpha = \pm} \left( \beta^2 \, \tau k_x - \alpha i k_y  \right) \, \Sigma_\alpha^{\,(1)} \, .
\end{equation}

Using Hamiltonian \eqref{AHam}, we arrive at eigenvalue equation  

\begin{equation}
\label{SpecEq}
\left[ 
\varepsilon^{\,(L)}_d ({\bf k}) \right]^3 - \left(
\frac{\beta \,c_0 \, \lambda_0}{2}
\right)^2 \, \varepsilon_d^{\,(L)} ({\bf k})
- \left( \hbar v_F \right)^2 \, 
\left\{ 
\left[ 
1 + \left( \frac{\beta \, \lambda_0}{2} \right)^2 \,
\right]^2 \, k_x^2 + 
\left[ 
1 + \left( \frac{\lambda_0}{2} \right)^2 \,
\right]^2 \, k_y^2
\right\} \, \varepsilon_d^{\,(L)} ({\bf k}) = 0 \, .
\end{equation}

We see that the dispersions demonstrate complete e-h symmetry and renormalized Fermi velocity. 
Even though the exact evaluation of the wave function is not possible even for ${ \bf k} = 0$
due to the complicated time dependence. 

\par 
\medskip 
\par 

This way, the energy two $\gamma = \pm 1$ dispersions still represent 
a Dirac cone with \textit{renormalized isotropic} Fermi velocity
\begin{equation}
 v_F^{\,(r)}(\lambda_0) =
v_F \left\{ 1 + \lambda_0^2/2 \left(1 + \lambda_0^2/8 \right) \right\}
\end{equation}
 
 \begin{equation}
\delta(\lambda_0, {\bf k}) = \frac{
\Delta_0(\beta=1, \lambda_0)
}{
\hbar \, v_F^{\,(r)}(\lambda_0) \, k
} \, .
 \end{equation}

 This quantity
is dimensionless, and it represents the induced energy gap related to a fixed electron energy for a 
given wave vector $k$.

\par 
\medskip 
\par 

Now we can obtain the wave functions, corresponding to dispersions \eqref{SpecRes} as the eigenstates of 
matrix \eqref{H3}. For the sake of simplicity, we will consider circularly-polarized light with $\beta = 1$.
The solutions, pertaining to the valence and conduction bands $\gamma = \pm 1$ are 

\begin{equation}
 \label{AAwf01}
 \Psi_d^{\gamma = \pm 1}(\tau\, \vert \, \lambda_0, k ) = 
 \frac{1}{\sqrt{\mc{N}^{(12)}}} \, \left\{ 
 \begin{array}{c} 
  \tau \, \mc{C}^{(1)} \, \tet{e}^{- i \tau \theta_{\bf k}}  \\
  \mc{C}^{(2)} \\
   \tau \, \tet{e}^{+ i \tau \theta_{\bf k}}  
 \end{array}
 \right\} \, ,
\end{equation}

where $\mc{C}^{(1)}(\tau \, \vert \, \lambda_0, k)$, $ \mc{C}^{(2)}(\tau \, \vert \, \lambda_0, 
k )$ and normalization factor $\mc{N}^{(12)}(\tau \, \vert \, \lambda_0,  k)$ are 

\begin{eqnarray}
\label{wf01}
 && \mc{C}^{(1)}(\gamma \, \vert \, \lambda_0, k) = 1 + 2 \, \delta^2(\lambda_0, k) +
 2 \, \gamma \, \delta(\lambda_0,  k) \, \sqrt{1 + \delta^2(\lambda_0, k)} \, ,  \\
 \nonumber 
 && \mc{C}^{(2)}(\gamma \, \vert \, \lambda_0, k) = \sqrt{2} \, \gamma \left[
 \sqrt{ 1 + \delta^2(\lambda_0, k)} 
  - \gamma \, \delta(\lambda_0, k)
 \right] \, , \\
 \nonumber 
 && \mc{N}^{\,(12)}(\gamma \, \vert \, \lambda_0, k) = 1 + 2 \,\left[ 
 \sqrt{1 + \delta^2(\lambda_0, k)} - \gamma \, \delta(\lambda_0, k)
 \right]^2 + \left\{
1 + 2 \delta(\lambda_0, k) \, \left[
\delta(\lambda_0, k) + \gamma \, \sqrt{1 + \delta^2(\lambda_0, k)} \,
\right] \,                                             
\right\}^2 \, . 
\end{eqnarray}

\par 

For the flat band with $\gamma = 0$, the correlative eigenstate is given as 

 \begin{equation}
\label{AAwf2}
\Psi_d^{\gamma = 0}(\tau\, \vert \, \lambda_0, {\bf k}) = \frac{1}{ \sqrt{
  2 \left[
 1 + \delta^2(\lambda_0, {\bf k})
 \, \right]
 }} \,
 \left\{  
 \begin{array}{c}
  \tet{e}^{- i \tau \theta_{ \bf k}}  \\
  \sqrt{2}\,\delta(\lambda_0, k) \\
  - \tet{e}^{+ i \tau \theta_{ \bf k}} 
 \end{array}
 \right\} \, .
\end{equation}

One can easily verify that the obtained wave functions \eqref{AAwf01} and \eqref{AAwf2} in the limit of 
vanishing electron-light interaction $\lambda_0 \rightarrow 0$ become equivalent to the dice lattice 
eigenstates, which are obtained from Eqs.~\eqref{Eig1} and \eqref{Eig2} at $\phi = \pi/4$.

\section{Linearly-polarized irradiation, derivation of Eqs.~\eqref{TA},   
\eqref{Twlfd}-\eqref{Twfld0} and \eqref{Twfl}-\eqref{Twfl0}.
}
\label{apb}

The total Hamiltonian of a quasiparticle interacting with linearly-polarized field is

\begin{equation}
\label{linham}
\hat{\mc{H}}({\bf k}, t) = \mbb{H}_\tau^{\phi}({\bf k})+\hat{\mbb{H}}_A^{(L)} \, .
\end{equation}

It acquires an additional interaction term 

\begin{equation}
 \hat{\mbb{H}}_A^{(L)} = - \tau \, c_0 \cos (\omega t) \, \left(
 \begin{array}{ccc}
  0 & \cos \phi & 0 \\
  \cos \phi  & 0 & \sin \phi \\
  0 & \sin \phi & 0 
 \end{array}
 \right) \, ,
\end{equation}
where the coupling amplitude $c_0 = v_F e E_0/\omega$ is identical to that in the case of 
elliptically- or circularly-polarized light.

\par 
So for $k_x=0$ and $k_y=0$ our equation is 

\begin{equation}
\label{Sch0}
 i \hbar \frac{d \psi_0(t)}{dt} = \hat{\mbb{H}}_A^{(L)} \psi_0 (t) \, .
\end{equation}

The solutions of Eq.~\eqref{Sch0} for the valence and conduction band edges with $\gamma = \pm 1$ are

\begin{equation}
 \psi_0^{\tau, \, \gamma} (t) = \frac{1}{\sqrt{2}} \,
 \left\{  
 \begin{array}{c}
  \tau \,\cos \phi \\
  \gamma \\
  \tau \, \sin \phi
 \end{array}
 \right\} \, \tet{exp}\left[ 
  i \gamma \, \lambda_0 \, \sin (\omega t)
 \right] \, ,
\end{equation}

and for the flat band

\begin{equation}
 \psi_0^{\gamma=0} (t) = 
 \left\{  
 \begin{array}{c}
  \sin \phi \\
  0 \\
  - \cos \phi
 \end{array}
 \right\} \, .
\end{equation}
As it appears, the last wave function has no time dependence so each part of Eq.~\eqref{Sch0} is equal to zero. 
These wave functions are clearly orthonormal and their structure demonstrates certain resemblance to the 
${\bf k}$-dependent eigenfunctions \eqref{Eig1} and \eqref{Eig2} for the $\alpha$-$T_3$ model. 
 
\par
\medskip
\par

The next step is to extend our solution to a finite wave vector. In order to achieve this, we need to solve 
the following time-dependent Schr{\H o}dinger equation

\begin{equation}
\label{Schk}
 i \hbar \, \frac{\pr }{\pr t} \Psi ({\bf k}, t) = \hat{\mc{H}}({\bf k}, t) \, \Psi ({\bf k}, t) 
\end{equation}
for complete Hamiltonian \eqref{linham}. At the $K$ point, this expression becomes identical to Eq.~\eqref{Sch0}.
We look for its solution in the form of the following expansion

\begin{equation}
\label{psik}
 \Psi ({\bf k}, t) = \sum\limits_{\gamma=-1}^{1} \mc{F}^{\,(\gamma)}({\bf k}, t) \, \psi_0^\gamma (t) \, ,
\end{equation}
in which the unknown time- and $k$-dependent coefficients $\mc{F}^{\,(\gamma)}({\bf k}, t)$ are to be found. 
Using Eq.~\eqref{Sch0} for ${\bf k} = 0$ and the orthogonality of its solutions $\psi_0^\gamma (t)$, our initial 
system \eqref{Schk} could be rewritten as 

\begin{equation}
\label{UTrans}
 i\hbar \, \frac{\pr}{\pr t} \mc{F}^{\,(\gamma)}({\bf k}, t)  = \sum\limits_{\rho = -1}^{1} \mc{F}^{\,(\rho)} \, 
 \langle \psi_0^\gamma (t) \, \vert
 \, \mbb{H}_\tau^{\phi}({\bf k}) \, \vert \, \psi_0^\rho (t) \rangle \, .
\end{equation}

This leads to a set of three coupled linear partial differential equations

\begin{eqnarray}
 && \frac{i}{v_F} \, \frac{\pr }{\pr t} \mc{F}^{\,(\mp 1)}({\bf k}, t) = \mp k_x \, \mc{F}^{\,( \mp 1)} \mp 
 \frac{i}{\sqrt{2}} \, \tet{e}^{ \pm 
 i z_\lambda(t)} \, \sin (2 \phi) \, k_y \, \mc{F}^{\,(0)} \mp 
 i \tet{e}^{ \pm 2 i z_\lambda(t)} \,\tau \cos (2 \phi) \, k_y  \, \mc{F}^{\,(\pm 1)} \, ,\\
 \nonumber 
 \text{and} \\
 \nonumber
 && \frac{\pr }{\pr t} \mc{F}^{\,(0)}({\bf k}, t) = 
 \frac{v_F}{\sqrt{2}} \,  \sin (2 \phi) \, k_y \, \sum\limits_{\alpha = \pm 1} \alpha \, \tet{e}^{ - i \alpha \, z_\lambda(t)} \,
 \mc{F}^{\,(-\alpha)} 
  \, ,
\end{eqnarray}
where $z_\lambda(t) = \lambda_0 \, \sin \omega t$. Using Floquet theorem, we look for the dressed state quasiparticle energy dispersions 
$\varepsilon_d ({\bf k})$ by means of the following substitution\,\cite{kisrep,ourJAP2017,hu}

\begin{equation}
\label{FloSub}
 \mc{F}^{(\gamma)} ({\bf k}, t) = \tet{exp}\left\{ 
 - \frac{i}{\hbar} \, \varepsilon_d ({\bf k}) \, t 
\right\}
\sum\limits_{n=-\infty}^{\infty} f^{\,(\gamma)}_n \tet{e}^{i n \omega t} \, .
\end{equation}
The second term in expression \eqref{FloSub} is a periodic function of time (with the period $T_0 = 2 \pi/\omega$), which is expanded in a 
Fourier series. Nested exponential function is traditionally reduced by the Jacobi-Anger series expansion 

\begin{equation}
 \tet{exp}\left\{ 
\pm \zeta \sin \omega t 
 \right\} = 
 \sum\limits_{m =-\infty}^{\infty} \mc{J}_m \left( \pm \zeta
 \right) \, \tet{e}^{i m \omega t} \, ,
\end{equation}

where $\mc{J}_m(\zeta)$ is the Bessel function of the first kind. We also apply the orthogonality condition 
of the Fourier expansion function, namely, 
\begin{equation}
 \nonumber 
 \int\limits_0^{2 \pi/ \omega} \tet{e}^{i \omega n \cdot t} \cdot \tet{e}^{- i \omega l \cdot t} \, dt = 
 \delta_{n,l} =  \begin{cases}
       1 \,\, \text{for} \,\, n=l \, ,\\
       0 \,\, \text{for} \,\, n \neq l
                  \end{cases}
\end{equation}
for any fixed number  $l$ $(- \infty < l < \infty)$. By doing so, we arrive at the following system of coupled 
linear algebraic equations 

\begin{eqnarray}
\nonumber
 && \left[  
 \mp k_x - \frac{\varepsilon_d ({\bf k})}{\hbar v_F} + l \omega
 \right] \, f^{\,(\mp 1)}_l \mp 
 i k_y \sum\limits_{m=-\infty}^{\infty} \left\{
 \frac{1}{\sqrt{2}} \,  \sin (2 \phi)  \, f^{\,(0)}_{l-m}
 \, \mc{J}_m \left( \pm \lambda_0 \right) +
 \tau \, \cos (2 \phi) \, f^{\,(\pm 1)}_{l-m} \mc{J}_m \left( \pm 2 \lambda_0 \right)
 \right\}  = 0  \\
 \nonumber 
 \text{and} \\
 && \left[  
- \frac{\varepsilon_d ({\bf k})}{\hbar v_F} + l \omega
 \right] \, f^{\,(0)}_l + \frac{i}{\sqrt{2}} \, \sin (2 \phi) \, k_y \sum\limits_{m=-\infty}^{\infty} \, 
 \sum\limits_{\alpha = \pm 1}
 \alpha \, \mc{J}_m \left(- \alpha \, \lambda_0 \right) \,  f^{\,(-\alpha)}_{l-m} 
 = 0 \, . 
\end{eqnarray}

Next, we follow the standard procedure\,\cite{kisrep, ourJAP2017} of eliminating the terms
$f^{\,(\rho)}_{l \neq 0} = 0$ $(\rho = -1,\,0,\,1)$. The point here is that for the case of 
off-resonant interaction, which satisfies $\hbar \omega \gg \varepsilon_d ({\bf k})$ and 
$\omega \gg v_F k_{\{x,y\}}$, all the terms in the square brackets in each equation of system
\eqref{nsyst} are negligible compared to $l \omega$ for any $l \neq 0$. The system is simplified 
to a new set of equations 

\begin{eqnarray}
\nonumber
\label{nsyst}
 &&  f^{\,(\mp 1)}_l =  \pm 
 \frac{i k_y}{l \omega} \sum\limits_{m=-\infty}^{\infty} \left\{
 \frac{1}{\sqrt{2}} \,  \sin (2 \phi)  \, f^{\,(0)}_{l-m}
 \, \mc{J}_m \left( \pm \lambda_0 \right) +
 \tau \, \cos (2 \phi) \, f^{\,(\pm 1)}_{l-m} \mc{J}_m \left( \pm 2 \lambda_0 \right)
 \right\}  \\
 \nonumber 
 \text{and} \\
 && f^{\,(0)}_l = - \frac{i}{\sqrt{2} \, l \omega} \, \sin (2 \phi) \, k_y \sum\limits_{m=-\infty}^{\infty} \, 
 \sum\limits_{\alpha = \pm 1}
 \alpha \, \mc{J}_m \left(- \alpha \, \lambda_0 \right) \,  f^{\,(-\alpha)}_{l-m}  \, ,
\end{eqnarray}

which can not be satisfied unless all the expansion coefficients $f^{\,(-1,0,1)}_l = 0$ for $l \neq 0$, at least
due to the fact that in this case one specific coefficient $f^{\,(\rho)}_l$ which we are looking for must be much 
smaller than all the others. Once only $l=0$ terms are left, our system \eqref{nsyst} is dramatically simplified 
and now becomes

\begin{equation}
\left[ 
 \tensor{\mc{K}}_{\,\tau} ({\bf k} \, \vert \, \phi, \lambda_0 \ll 1) 
  - \frac{\varepsilon_d ({\bf k })}{\hbar v_F}
  \right] \otimes \left\{
 \begin{array}{c}
  f^{\,(-1)}_0 \\
  f^{\,(0)}_0 \\
  f^{\,(1)}_0
 \end{array}
\right\} = 0 \, ,
\end{equation}

where 
\begin{eqnarray}
\label{finalHam}
&& \tensor{\mc{K}}_{\,\tau} ({\bf k} \, \vert \, \phi, \lambda_0 \ll 1)  = 
\left[ 
\begin{array}{ccc}
 - k_x/2
 & - i/\sqrt{2} \, \sin (2 \phi) \, k_y \, \mc{J}_0 (\lambda_0) 
 & - i \tau \,  \cos (2 \phi) \, k_y \, \mc{J}_0 (2 \lambda_0) \\
 %%%
0 & 0 &
 - i/\sqrt{2} \, \sin (2 \phi) \, k_y \,  \mc{J}_0 (\lambda_0) \\
 %%%
0 & 0 & k_x/2
                                                           \end{array}
\right] \,\, + \,\, h.\,c. \, \,\,\, .
\end{eqnarray}
Here, $h.\,c.$ means adding a Hermitian conjugate matrix, exactly as we had in Eq.~\eqref{H3}, and we also used 
the fact that zero-order Bessel function is even, i.e., $\mc{J}_0 (\xi) = \mc{J}_0 (- \xi) \backsimeq 1 - \xi^2/4$,
 $\xi = \lambda_0$ and $2 \lambda_0$.

\par 
As the final result of this derivation, we obtain the dressed state quasiparticle energy dispersions 

\begin{eqnarray}
 && \varepsilon_d ({\bf k}) = 0 \,\,\, \text{and} \\
 \nonumber 
 && \varepsilon_d ({\bf k}) = \pm \hbar v_F  \, \sqrt{\mbb{A}(\theta_{\bf k} \, \vert \, \phi, \lambda_0)} \, k \, ,
\end{eqnarray}

which angular dependence 

\begin{equation}
\label{A}
\mbb{A}(\theta_{\bf k} \, \vert \, \phi, \lambda_0) = 
 \cos^2 \theta + \left\{
 \left[
 \mc{J}_0 (2 \lambda_0) \, \cos (2 \phi)
 \right]^2 + 
 \left[
 \mc{J}_0 (\lambda_0) \, \sin (2 \phi)
 \right]^2
 \right\} \,\, \sin^2 \theta
\end{equation}

clearly demonstrates the anisotropy of the energy dispersions induced by the electron-photon interaction. We immediately 
recover the corresponding result graphene\,\cite{kisrep} if $\phi$ is equal to zero. For a dice lattice with $\phi = \pi/4$,  
only second term $\mc{J}_0 (\lambda_0) \, \sin (2 \phi)$ remains non-zero and the effect of dressing field is the lowest.

\par 
In our case of off-resonant radiation with low intensity with $\lambda = c_0^{(L)}/ (\hbar \omega) \ll 1$, 
the zero-order Bessel function of the first kind could be expanded as shown above, so that the conduction and 
valence bands energy dispersions are further approximated as 

\begin{equation} 
 \varepsilon_d^{\gamma = \pm 1} ({\bf k}) \backsimeq \gamma \hbar v_F \, \left\{ k^2  -
 \frac{\lambda_0^2}{4} \, \left[ 5 + 3 \cos (4 \phi) \right]
\, k_y^2 
\right\}^{1/2} \, .
\end{equation}

The anisotropy and the difference between the  Fermi velocities in the $k_x$ and $k_y$ directions 
in both valence and conduction bands is the largest for $\phi = 0$ and the smallest for a dice 
lattice case. 

\par 
\medskip
\par 

The corresponding wave function for the dressed state quisiparticle, obtained directly from Eq.~\eqref{psik},
is

\begin{eqnarray}
\label{wfl}
 && \Psi_d^{\gamma = \pm 1} (\lambda_0, {\bf k}) =  
\frac{1}{
\left[
 \mc{N}_{23}^{\,(\phi)}
 \right]^{1/2}
}
\, \times \,
 \tet{exp}\left[ \mp i v_F k t \, f_\theta^{\,(\phi)}
 \right] \, \times
 \\ 
 \nonumber 
&& \times \, \left[
 %% T1
 \frac{ r_1^{\,(\phi)}}{\sqrt{2}} \,
 \left\{  
 \begin{array}{c}
  \tau \, \cos \phi\\
  \pm 1\\
  \tau \, \sin \phi
 \end{array}
 \right\} \,  \tet{e}^{\pm i z_\lambda(t)} + 
% T2
 r_2^{\,(\phi)} \,
\left\{  
 \begin{array}{c}
  \sin \phi \\
  0 \\
  - \cos \phi
 \end{array}
 \right\} +
 % T3
\frac{r_3^{\,(\phi)}}{\sqrt{2}} \,
 \left\{  
 \begin{array}{c}
  \tau \cos \phi\\
  \mp 1 \\
  \tau \sin \phi
 \end{array}
 \right\} \, \tet{e}^{ \mp i z_\lambda (t)}
 \right] \, ,
\end{eqnarray}

where

\begin{eqnarray}
 && r_1^{\,(\phi)} (\lambda_0, {\bf k}) = 2 f_\theta^{\,(\phi)} \left(f_\theta^{\,(\phi)}+ \cos \theta_{\bf k}
 \right) - \mc{J}_0^2(\lambda_0) \sin^2 \theta_{\bf k} \sin^2 (2 \phi) \, , \\
 \nonumber 
 && r_2^{\,(\phi)} (\lambda_0, {\bf k}) = - i \sqrt{2} \,
\sin \theta_{\bf k} \, \sin (2 \phi) \,  \mc{J}_0(\lambda_0)  \, \left[ 
 f_\theta^{\,(\phi)} + \cos \theta_{\bf k} + i \tau \, \sin \theta_{\bf k} \, \cos (2 \phi)  \,
 \mc{J}(2\lambda_0) \, \right] \, ,\\
 \nonumber 
 && r_{3}^{\,(\phi)} (\lambda_0, {\bf k}) = - 
  \sin \theta_{\bf k}  \left[ 
  \sin \theta_{\bf k} \, \sin^2 (2 \phi) \, \mc{J}_0^2(\lambda_0) +
  2 i \tau \, f_\theta^{\,(\phi)} \, \cos (2 \phi) \, \mc{J}_0 (2\lambda_0)  
  \right] \, .
  \end{eqnarray}
  
  The normalization factor is done by evaluating each component of the wave function
  \begin{eqnarray}
 && \mc{N}_{23}^{\,(\phi)} = \left| \frac{\tau \, \cos \phi}{\sqrt{2}} \, \left[ 
  r_1^{\,(\phi)} (\lambda_0, {\bf k}) +  r_3^{\,(\phi)} (\lambda_0, {\bf k})
 \right] \, + r_2^{\,(\phi)} (\lambda_0, {\bf k}) \, \sin \phi \, 
 \right|^2 + \frac{1}{2} \,\left|  r_1^{\,(\phi)} (\lambda_0, {\bf k}) - \right. 
 \\
 \nonumber
 && - \left. r_3^{\,(\phi)} (\lambda_0, {\bf k}) \right|^2 +  
 \left| \frac{\tau \, \sin \phi}{\sqrt{2}} \, \left[ 
  r_1^{\,(\phi)} (\lambda_0, {\bf k}) +  r_3^{\,(\phi)} (\lambda_0, {\bf k})
 \right] \, - r_2^{\,(\phi)} (\lambda_0, {\bf k}) \, \cos \phi \, 
 \right|^2 \, . 
\end{eqnarray}

\par 
\medskip 
\par 

For the flat band with $\gamma = 0$, the wave function takes the form

\begin{equation}
\label{wfl0}
 \Psi_d^{0} (\lambda_0, {\bf k}) = \frac{1}{ 
 \sqrt{2 \, \mc{N}_{\theta}^{\,(\phi)}} }
 \, \times \, 
\left[ 
 r_1^{\,(\phi)} \, \frac{\tau}{\sqrt{2}} \, \sum\limits_{\alpha = \pm 1}
 \tet{e}^{i \alpha \, z_\lambda(t)} \,
 \left\{  
 \begin{array}{c}
  \cos \phi \\
  \alpha \, \tau  \\
  \sin \phi
 \end{array}
\right\}  + r_2^{\,(\phi)} \, 
\left\{  
 \begin{array}{c}
  \sin \phi \\
  0 \\
  - \cos \phi
 \end{array}
 \right\}
\, \right] \, , 
\end{equation}

where 

\begin{eqnarray}
\label{rsf}
 && r_1^{\,(\phi)}(\lambda_0, {\bf k}) = - i \sin (2 \phi) \, \sin \theta_{\bf k} \, \mc{J}_0 (\lambda_0) \, , \\
 \nonumber 
 && r_2^{\,(\phi)}(\lambda_0, {\bf k}) =  \sqrt{2} \, \left[ 
 \cos  \theta_{\bf k} + i \tau \, \cos (2 \phi) \, \sin \theta_{\bf k} \, \mc{J}_0 (2\lambda_0)
 \right] \, .
 \nonumber
\end{eqnarray}

or, clearly presenting its component, we write 

\begin{eqnarray}
 && \Psi_d^{0} (\lambda_0, {\bf k}) = \frac{1}{\sqrt{\mc{N}_{\theta}^{\,(\phi)}}} \,  
\left\{ 
\begin{array}{c}
 \sin \phi \, \left[
 \cos \theta_{\bf k} - i \tau \, \sin \theta_{\bf k} \, \mbb{X}^{\,(\phi)}_\theta (\lambda_0) \,
 \right] \\
 0 \\
 - \cos \phi \, \left[
 \cos \theta_{\bf k} + i \tau \, \sin \theta_{\bf k} \, 
 \left( 2 \, \mc{J}_0 (\lambda_0) -
 \mbb{X}^{\,(\phi)}_\theta (\lambda_0) 
 \, \right) \,
 \right]
\end{array}
\right\} \, , \\
\nonumber 
&& \mbb{X}^{\,(\phi)}_\theta (\lambda_0) = 2 \cos^2 \phi \,
\mc{J}_0 (\lambda_0) - \cos (2 \phi) \, \mc{J}_0 (2 \lambda_0) \backsimeq 1 - \frac{\lambda_0^2}{4} \,
\left[ 4 - 3 \cos (2 \phi) \right] \, , \\
\nonumber 
&&  2 \, \mc{J}_0 (\lambda_0) -
 \mbb{X}^{\,(\phi)}_\theta (\lambda_0) \backsimeq  1 - \frac{\lambda_0^2}{4} \,
\left[4 + 3 \cos (2 \phi) \right] \, , \\
\nonumber
&& \mc{N}_{\theta}^{\,(\phi)} = \cos^2 \theta_{\bf k} + \sin^2 \theta_{\bf k}
\left\{ \, \left[ 
 \, \mbb{X}^{\,(\phi)}_\theta (\lambda_0) \right]^2 
 + \left( 2 \cos \phi \right)^2 \,
 \mc{J}_0 (\lambda_0) \, \left[
\mc{J}_0 (\lambda_0) - \mbb{X}^{\,(\phi)}_\theta (\lambda_0)
\right] \, 
\right\}
\backsimeq \\
\nonumber 
&& \backsimeq 1 - \frac{\lambda_0^2}{8} \left\{
5 - \left[
5 \cos (2 \, \theta_{\bf k}) - 6 \cos (4 \phi) \, \sin^2 \theta_{\bf k}
\right] \,
\right\} \, .
\end{eqnarray}

\par 
\medskip 
\par 

\subsection{Dice lattice, $\alpha = 1.0$}

As a general rule, the results are greatly simplified for the case of a dice lattice

\begin{eqnarray}
\label{wfld}
 && \Psi_d^{\gamma = \pm 1} (\lambda_0, {\bf k}) = \tet{exp}[ \mp i v_F k t \, f_\theta] \, 
 \frac{f_\theta + \cos\theta_{\bf k}}{4 \,f_\theta} \, \times \,
  \left[
 %% T1
 \tet{e}^{\pm i z_\lambda(t)} \,
 \left\{  
 \begin{array}{c}
  \tau \\
  \pm \sqrt{2} \\
  \tau
 \end{array}
 \right\} - \right. \\ 
 \nonumber 
&& \left.
% T2
  - \, 2 i \, \frac{\mc{J}_0(\lambda_0) \, \sin \theta_{\bf k}}{f_\theta + \cos \theta_{\bf k}} \,
\left\{  
 \begin{array}{c}
  1 \\
  0 \\
  - 1
 \end{array}
 \right\} -
 % T3
 \left[ 
 \frac{\mc{J}_0(\lambda_0) \, \sin \theta_{\bf k} }{f_\theta + \cos \theta_{\bf k}}
 \right]^2
 \left\{  
 \begin{array}{c}
  \tau \\
  \mp \sqrt{2} \\
  \tau
 \end{array}
 \right\} \, \tet{e}^{ \mp i z_\lambda (t)}
 \right] \, ,
\end{eqnarray}

where 

\begin{equation}
\label{drf}
 f_\theta = \sqrt{\,\mbb{A}(\theta_{\bf k} \, \vert \, \phi = \pi/4, \lambda_0) } =
 \left\{ 
 \cos^2 \theta + \left[
 \sin \theta \,
 \mc{J}_0 (\lambda_0) \,
 \right 
 ]^2 \,
 \right\}^{1/2}
\end{equation}

and we once again use our previous notation $z_\lambda(t) = \lambda_0 \, \sin (\omega t)$. 
This wave function could be presented in a specifically simplified way 

\begin{eqnarray}
 && \Psi_d^{\gamma = \pm 1} (\lambda_0, {\bf k}) = \frac{1}{2} \, 
 \left\{
 \begin{array}{c}
  \tau \, \tet{e}^{-i \Phi_\theta (\lambda_0)} \\
  \gamma \, \sqrt{2} \\
  \tau \, \tet{e}^{i \Phi_\theta (\lambda_0)}
 \end{array}
 \right\} \, , \\
 \nonumber 
 && \Phi_\theta (\lambda_0) = 2 \arctan \left[ \tau \,
 \frac{\mc{J}_0(\lambda_0)\,\sin \theta_{\bf k}}{f_\theta + \cos \theta_{\bf k}} 
 \right] \backsimeq \tau \, \left[ \theta_{\bf k} - \frac{\lambda_0^2}{8} \, 
 \sin^2 (2  \theta_{\bf k}) + ... \right] \, .
\end{eqnarray}

Phase $\Phi_\theta (\lambda_0)$ for a dice lattices differs from that in graphene by the 
argument of the Bessel function and the expansion coefficient. 

\par 

The remaining wave 
function for the flat band with $\gamma = 0$ takes the following form

\begin{equation}
\label{wfld0}
 \Psi_d^{0} (\lambda_0, {\bf k}) = \frac{1}{\sqrt{2} \,f_\theta} \, \times \, 
\left[ 
 - \frac{i \tau}{2} \, \sin \theta_{\bf k} \, \mc{J}_0(\lambda_0) \,
  \sum\limits_{\alpha = \pm 1}
 \tet{e}^{i \alpha \, z_\lambda(t)} \,
 \left\{  
 \begin{array}{c}
  1 \\
  \alpha \sqrt{2} \, \tau  \\
  1
 \end{array}
\right\}  + \cos \theta_{\bf k} \, 
\left\{  
 \begin{array}{c}
  1 \\
  0 \\
  - 1
 \end{array}
 \right\}
\, \right] \, .
\end{equation}

The structure of the obtained wave function \eqref{wfld0} is such that at $t=0$ it consists 
of two components of equal amplitudes and phase difference $\Phi_\theta (\lambda_0)$,
and could be rewritten as 

\begin{eqnarray}
&& \Psi_d^{0} (\lambda_0, {\bf k}) = \frac{1}{\sqrt{2}} \, 
\left\{ 
\begin{array}{c}
 \tet{e}^{- i \Phi_\theta (\lambda_0)} \\
 0 \\
 - \tet{e}^{i \Phi_\theta (\lambda_0)}
\end{array}
\right\} \, , \\
\nonumber 
&& \Phi_\theta (\lambda_0) = \arctan \left[ 
\tau \, \mc{J}_0(\lambda_0) \, \tan \theta_{\bf k}
\right] \backsimeq \tau \left[
\theta_{\bf k} - \frac{\lambda^2}{2} \, \sin (2 \theta_{\bf k}) + ...
\right] \, .
\end{eqnarray}

It is straightforward to verify that in the limit of vanishing electron-photon interaction $\lambda_0 = c_0/(\hbar
\omega) \Longrightarrow 0$ all our obtained wave functions \eqref{wfl}-\eqref{wfl0} and \eqref{wfld}-\eqref{wfld0} 
are exactly equivalent to the $\alpha$-T$_3$ lattice eigenstates \eqref{Eig1} and \eqref{Eig2} or 
to the corresponding dice lattice limit with $\phi = \pi/4$.

%%%%%%%%%%%%%%%%%%%%%%%%%%%%%%%
%%%%%%%%%%%%%%%%%%%%%%%%%%%%%%%%%%%%%

\section{Dressing field-induced adjustments to Berry phases, derivation of Eq.~\eqref{TB1} and 
theoretical background for our numerical results presented in Fig.~\ref{FIG:6}.
}
\label{apc}

In this appendix, we provide the details of the \textit{Berry phase} evaluation for the obtained 
dresses states wave functions \eqref{Twf1} and \eqref{Twf2} for the case of circularly-polarized 
field applied to a dice lattice, as well as for eigenstates \eqref{Twlfd}-\eqref{Twfld0} and 
\eqref{Twfl}-\eqref{Twfl0}, corresponding to various types of $\alpha$-T$_3$ lattices interacting 
with irradiation with linear polarization. 

\par
\medskip 
\par 

In polar coordinates, the general expression for gradient $\vec{\nabla}_k$ and 
the vector length element $d{\bf k}$\,\cite{arfken}

\begin{eqnarray}
 && \vec{\nabla}_k = \frac{\pr}{\pr k}\, \hat{e}_k + \frac{1}{k} 
 \frac{\pr}{\pr \theta_{\bf k}} \, \hat{e}_\theta \, , \\
 \nonumber 
 && d {\bf k} = dk \, \hat{e}_k + k \, d\theta_{\bf k} \,\, \hat{e}_\theta \, .
\end{eqnarray}

As the first step, we must choose the right \textit{closed integration path} for Eq.~\eqref{Berry}.
In order to satisfy the requirement of an adiabatic, or isoenergetic evolution of a quantum system, 
during which a Berry phase is acquired, we have to choose a path with a constant energy of our quisparticle,
i.e., with $\varepsilon^{\gamma = \pm 1}_{\tau, \, \phi}({\bf k}) = \varepsilon_0 = const$.

For the first case of a circularly-polarized light, the energy dispersions \eqref{SpecRes} and the corresponding 
eigenstates \eqref{Twf1} and \eqref{Twf2} are isotropic, so that the required path ought to be 
\textit{a circle of radius} $k^{\,(0)}$. While the wave function components and their scalar product 
can still depend on both $k-$ and $\theta_{\bf k}$-components of the two-dimensional vector ${\bf k}$, 
the integration variable in Eq.~\eqref{Berry} is modified as $d {\bf k} = k^{\,(0)} \, d\theta_{\bf k} \, 
\hat{e}_\theta$.

\par 
We begin with the eigenstate \eqref{Twf1} for the valence and conduction bands. 
In the simplest case of a circular path of radius $k^{\,(0)}$,
the Berry phase defined in Eq.~\eqref{Berry} is equal to

\begin{equation}
 \phi^{\,(e)}_B (\tau = 1) = \frac{
 \left[\mc{C}^{(1)}(k^{\,(0)}) \right]^2 - 1
 }{
 \mc{N}^{\,(e)}(k^{\,(0)})
 } \,
 \int\limits_{0}^{2 \pi} d \theta_{\bf k}  \backsimeq \pi \, \gamma
 \, \frac{c_0}{\hbar v_F k^{\,(0)}} \, \lambda_0 \, .
\end{equation}

while in the vicinity of $K'$ valley this result is exactly opposite $\phi^{\,(e)}_B (\tau = - 1) =
- \phi^{\,(e)}_B (\tau = 1) $, similarly to the case of 
non-irradiated $\alpha$-T$_3$. In the absence of the irradiation, the Berry phase for a dice 
lattice is zero - $\phi^{\,(0)}_B (\tau) = \tau \pi \, \cos(\pi/2) = 0$.\,\cite{A1Thesis} The 
corresponding Berry phase for the flat band remains zero in the presence of circularly-polarized
light, which could be confirmed by evaluating integral \eqref{Berry} for wave function \eqref{Twf2}. 

\par
\medskip
\par 

In contrast, the constant-energy cut for dispersions \eqref{linD} with angular dependence \eqref{TA}
has an elliptic shape, as it shown in Fig.~\ref{FIG:4} $(a)$. Such an ellipse is described as

\begin{equation}
 \left(\frac{k_x^{\,(e)}}{a} \right)^2 + \left[\frac{k_y^{\,(e)}}{b(\lambda_0, \phi)} \right]^2 = 1 \, , 
\end{equation}

where 

\begin{eqnarray}
 && a = \frac{\varepsilon_0}{\hbar v_F} \, , \\
 \nonumber 
 && b(\lambda_0, \phi) =  \frac{\varepsilon_0}{\hbar v_F} \, \left\{
 \left[
 \mc{J}_0 (2 \lambda_0) \, \cos (2 \phi)
 \right]^2 + 
 \left[
 \mc{J}_0 (\lambda_0) \, \sin (2 \phi)
 \right]^2 \, 
 \right\}^{-1} \backsimeq a \,\, 
 \left\{ 1  + \frac{\lambda_0^2}{8} \, \left[ 5 + 3 \cos (4 \phi) \right]
\right\} > a \, .
\end{eqnarray}

Elliptical path in polar coordinates is given by

\begin{eqnarray}
 && k^{\,(e)}(\theta_{\bf k}) = a b(\lambda_0, \phi)
 \, \left\{
 \left(a \, \cos \theta_{\bf k} \right)^2 +
 \left[ b(\lambda_0, \phi) \, \sin \theta_{\bf k} \right]^2
 \right \}^{-1/2} =
 \frac{b}{\sqrt{
 1- \left[ e(\lambda_0, \phi) \, \cos \theta_{\bf k} \right]^2
 }} \, ,
\end{eqnarray}
where $a$ and $b(\lambda_0, \phi)$ are major and minor semi-axes of an ellipse in the $k$-space and
$e(\lambda_0) = \sqrt{1- \left[ b(\lambda_0, \phi)/a \right]^2}$
is its \textit{eccentricity}. We see that each specific elliptical path and its eccentricity depends on the
incoming radiation intensity or coupling constant $\lambda_0$. Our notation $\theta_{\bf k} = \arctan (k_y/k_x
)$ should not lead to a confusion since in the polar coordinates angle $\theta_{\bf k}$ is independent from
the radial component $k$.

\begin{equation}
 d k^{\,(e)}(\theta_{\bf k}) = - \frac{1}{2} \, b(\lambda_0, \phi) e(\lambda_0, \phi)^2 \,
 \frac{\sin(2 \theta_{\bf k}) \, 
 d \theta_{\bf k}}{\left[
 1 - e(\lambda_0, \phi)^2 \, \cos \theta_{\bf k}
 \right]^{2/3}
 } 
\end{equation}
and 
\begin{equation}
 d {\bf k} = \frac{b(\lambda_0, \phi) \, d \theta_{\bf k} }{\sqrt{1 - e(\lambda_0, \phi)^2 \, 
 \cos \theta_{\bf k}}} \, \left\{
   \hat{e}_\theta 
 - \frac{e(\lambda_0, \phi)^2 \, \sin(2 \theta_{\bf k})}{2 \,\left[
 1 - e(\lambda_0, \phi)^2 \, \cos \theta_{\bf k}
 \right]
 }
  \, \hat{e}_k  
 \right\} \, .
\end{equation}

\par 
\medskip
\par 

In the case of linearly-polarized dressing field, the wave function components depend only on angle 
$\theta_{\bf k}$ but not on the radial component $k$ for all possible values of $\alpha$. 

For $\alpha = 1$, all the relevant $\hat{e}_{\theta}$-components of the gradient all 
equal to zero. Indeed, for wave functions \eqref{new1} or \eqref{new2},

\begin{equation}
\vec{\nabla}_{{\bf k}}
 \Psi_{\tau,\,\phi}^{\gamma}({\bf k}) = \frac{1}{k^{\,(e)}(\theta_{\bf k})} \, \frac{\pr}{\pr \theta_{\bf
 k}} \, \Psi_{\tau,\,\phi}^{\gamma}({\bf k}) = \frac{i \tau}{2 k^{\,(e)}(\theta_{\bf k})} \,
 \tet{e}^{-i \Phi_\theta (\lambda_0)} \, \frac{\pr \Phi_\theta (\lambda_0) }{ \pr 
  \theta_{\bf k}} \,
 \left\{
 \begin{array}{c}
  - 1  \\
  0 \\
  1
 \end{array}
 \right\} \,
\end{equation}

so that $\left[ \Psi_{\tau,\,\phi}^{\gamma}({\bf k}) \right]^\dag \vec{\nabla}_{{\bf k}}
 \Psi_{\tau,\,\phi}^{\gamma}({\bf k})=0$.
 
The only difference between the valence/conduction and the flat bands here is the exact expression 
of $\Phi_\theta (\lambda_0)$, which does not play any crucial role. The middle 
component of each wave function does not contribute to Eq.~\eqref{Berry} since its $\pr/ \pr \theta_{\bf 
k}$ derivative is always zero.

\bibliography{DS-AT3}

\begin{thebibliography}{53}
\expandafter\ifx\csname natexlab\endcsname\relax\def\natexlab#1{#1}\fi
\expandafter\ifx\csname bibnamefont\endcsname\relax
  \def\bibnamefont#1{#1}\fi
\expandafter\ifx\csname bibfnamefont\endcsname\relax
  \def\bibfnamefont#1{#1}\fi
\expandafter\ifx\csname citenamefont\endcsname\relax
  \def\citenamefont#1{#1}\fi
\expandafter\ifx\csname url\endcsname\relax
  \def\url#1{\texttt{#1}}\fi
\expandafter\ifx\csname urlprefix\endcsname\relax\def\urlprefix{URL }\fi
\providecommand{\bibinfo}[2]{#2}
\providecommand{\eprint}[2][]{\url{#2}}

\bibitem[{\citenamefont{Malcolm and Nicol}(2016)}]{malcolmMain}
\bibinfo{author}{\bibfnamefont{J.~D.} \bibnamefont{Malcolm}} \bibnamefont{and}
  \bibinfo{author}{\bibfnamefont{E.~J.} \bibnamefont{Nicol}},
  \bibinfo{journal}{Physical Review B} \textbf{\bibinfo{volume}{93}},
  \bibinfo{pages}{165433} (\bibinfo{year}{2016}).

\bibitem[{\citenamefont{Dey and Ghosh}(2018)}]{hu}
\bibinfo{author}{\bibfnamefont{B.}~\bibnamefont{Dey}} \bibnamefont{and}
  \bibinfo{author}{\bibfnamefont{T.~K.} \bibnamefont{Ghosh}},
  \bibinfo{journal}{Physical Review B} \textbf{\bibinfo{volume}{98}},
  \bibinfo{pages}{075422} (\bibinfo{year}{2018}).

\bibitem[{\citenamefont{Novoselov et~al.}(2005)\citenamefont{Novoselov, Geim,
  Morozov, Jiang, Katsnelson, Grigorieva, Dubonos, and Firsov}}]{gr01}
\bibinfo{author}{\bibfnamefont{K.}~\bibnamefont{Novoselov}},
  \bibinfo{author}{\bibfnamefont{A.~K.} \bibnamefont{Geim}},
  \bibinfo{author}{\bibfnamefont{S.}~\bibnamefont{Morozov}},
  \bibinfo{author}{\bibfnamefont{D.}~\bibnamefont{Jiang}},
  \bibinfo{author}{\bibfnamefont{M.}~\bibnamefont{Katsnelson}},
  \bibinfo{author}{\bibfnamefont{I.}~\bibnamefont{Grigorieva}},
  \bibinfo{author}{\bibfnamefont{S.}~\bibnamefont{Dubonos}}, \bibnamefont{and}
  \bibinfo{author}{\bibfnamefont{A.}~\bibnamefont{Firsov}},
  \bibinfo{journal}{nature} \textbf{\bibinfo{volume}{438}},
  \bibinfo{pages}{197} (\bibinfo{year}{2005}).

\bibitem[{\citenamefont{Nair et~al.}(2008)\citenamefont{Nair, Blake,
  Grigorenko, Novoselov, Booth, Stauber, Peres, and Geim}}]{gr02}
\bibinfo{author}{\bibfnamefont{R.~R.} \bibnamefont{Nair}},
  \bibinfo{author}{\bibfnamefont{P.}~\bibnamefont{Blake}},
  \bibinfo{author}{\bibfnamefont{A.~N.} \bibnamefont{Grigorenko}},
  \bibinfo{author}{\bibfnamefont{K.~S.} \bibnamefont{Novoselov}},
  \bibinfo{author}{\bibfnamefont{T.~J.} \bibnamefont{Booth}},
  \bibinfo{author}{\bibfnamefont{T.}~\bibnamefont{Stauber}},
  \bibinfo{author}{\bibfnamefont{N.~M.} \bibnamefont{Peres}}, \bibnamefont{and}
  \bibinfo{author}{\bibfnamefont{A.~K.} \bibnamefont{Geim}},
  \bibinfo{journal}{Science} \textbf{\bibinfo{volume}{320}},
  \bibinfo{pages}{1308} (\bibinfo{year}{2008}).

\bibitem[{\citenamefont{Ferrari et~al.}(2006)\citenamefont{Ferrari, Meyer,
  Scardaci, Casiraghi, Lazzeri, Mauri, Piscanec, Jiang, Novoselov, Roth
  et~al.}}]{gr03}
\bibinfo{author}{\bibfnamefont{A.~C.} \bibnamefont{Ferrari}},
  \bibinfo{author}{\bibfnamefont{J.}~\bibnamefont{Meyer}},
  \bibinfo{author}{\bibfnamefont{V.}~\bibnamefont{Scardaci}},
  \bibinfo{author}{\bibfnamefont{C.}~\bibnamefont{Casiraghi}},
  \bibinfo{author}{\bibfnamefont{M.}~\bibnamefont{Lazzeri}},
  \bibinfo{author}{\bibfnamefont{F.}~\bibnamefont{Mauri}},
  \bibinfo{author}{\bibfnamefont{S.}~\bibnamefont{Piscanec}},
  \bibinfo{author}{\bibfnamefont{D.}~\bibnamefont{Jiang}},
  \bibinfo{author}{\bibfnamefont{K.}~\bibnamefont{Novoselov}},
  \bibinfo{author}{\bibfnamefont{S.}~\bibnamefont{Roth}}, \bibnamefont{et~al.},
  \bibinfo{journal}{Physical review letters} \textbf{\bibinfo{volume}{97}},
  \bibinfo{pages}{187401} (\bibinfo{year}{2006}).

\bibitem[{\citenamefont{Rizzi et~al.}(2006)\citenamefont{Rizzi, Cataudella, and
  Fazio}}]{at2}
\bibinfo{author}{\bibfnamefont{M.}~\bibnamefont{Rizzi}},
  \bibinfo{author}{\bibfnamefont{V.}~\bibnamefont{Cataudella}},
  \bibnamefont{and} \bibinfo{author}{\bibfnamefont{R.}~\bibnamefont{Fazio}},
  \bibinfo{journal}{Physical Review B} \textbf{\bibinfo{volume}{73}},
  \bibinfo{pages}{144511} (\bibinfo{year}{2006}).

\bibitem[{\citenamefont{Vidal et~al.}(1998)\citenamefont{Vidal, Mosseri, and
  Dou{\c{c}}ot}}]{at1}
\bibinfo{author}{\bibfnamefont{J.}~\bibnamefont{Vidal}},
  \bibinfo{author}{\bibfnamefont{R.}~\bibnamefont{Mosseri}}, \bibnamefont{and}
  \bibinfo{author}{\bibfnamefont{B.}~\bibnamefont{Dou{\c{c}}ot}},
  \bibinfo{journal}{Physical review letters} \textbf{\bibinfo{volume}{81}},
  \bibinfo{pages}{5888} (\bibinfo{year}{1998}).

\bibitem[{\citenamefont{Goldman and Dalibard}(2014)}]{prX1}
\bibinfo{author}{\bibfnamefont{N.}~\bibnamefont{Goldman}} \bibnamefont{and}
  \bibinfo{author}{\bibfnamefont{J.}~\bibnamefont{Dalibard}},
  \bibinfo{journal}{Physical review X} \textbf{\bibinfo{volume}{4}},
  \bibinfo{pages}{031027} (\bibinfo{year}{2014}).

\bibitem[{\citenamefont{Perez-Piskunow
  et~al.}(2014)\citenamefont{Perez-Piskunow, Usaj, Balseiro, and Torres}}]{p1}
\bibinfo{author}{\bibfnamefont{P.}~\bibnamefont{Perez-Piskunow}},
  \bibinfo{author}{\bibfnamefont{G.}~\bibnamefont{Usaj}},
  \bibinfo{author}{\bibfnamefont{C.}~\bibnamefont{Balseiro}}, \bibnamefont{and}
  \bibinfo{author}{\bibfnamefont{L.~F.} \bibnamefont{Torres}},
  \bibinfo{journal}{Physical Review B} \textbf{\bibinfo{volume}{89}},
  \bibinfo{pages}{121401} (\bibinfo{year}{2014}).

\bibitem[{\citenamefont{Morell and Torres}(2012)}]{p3}
\bibinfo{author}{\bibfnamefont{E.~S.} \bibnamefont{Morell}} \bibnamefont{and}
  \bibinfo{author}{\bibfnamefont{L.~E.~F.} \bibnamefont{Torres}},
  \bibinfo{journal}{Physical Review B} \textbf{\bibinfo{volume}{86}},
  \bibinfo{pages}{125449} (\bibinfo{year}{2012}).

\bibitem[{\citenamefont{Gu et~al.}(2011)\citenamefont{Gu, Fertig, Arovas, and
  Auerbach}}]{p4}
\bibinfo{author}{\bibfnamefont{Z.}~\bibnamefont{Gu}},
  \bibinfo{author}{\bibfnamefont{H.}~\bibnamefont{Fertig}},
  \bibinfo{author}{\bibfnamefont{D.~P.} \bibnamefont{Arovas}},
  \bibnamefont{and} \bibinfo{author}{\bibfnamefont{A.}~\bibnamefont{Auerbach}},
  \bibinfo{journal}{Physical review letters} \textbf{\bibinfo{volume}{107}},
  \bibinfo{pages}{216601} (\bibinfo{year}{2011}).

\bibitem[{\citenamefont{Kozin et~al.}(2018)\citenamefont{Kozin, Iorsh, Kibis,
  and Shelykh}}]{koz1}
\bibinfo{author}{\bibfnamefont{V.}~\bibnamefont{Kozin}},
  \bibinfo{author}{\bibfnamefont{I.}~\bibnamefont{Iorsh}},
  \bibinfo{author}{\bibfnamefont{O.}~\bibnamefont{Kibis}}, \bibnamefont{and}
  \bibinfo{author}{\bibfnamefont{I.}~\bibnamefont{Shelykh}},
  \bibinfo{journal}{Physical Review B} \textbf{\bibinfo{volume}{97}},
  \bibinfo{pages}{035416} (\bibinfo{year}{2018}).

\bibitem[{\citenamefont{Mandal et~al.}(2018)\citenamefont{Mandal, Liew, and
  Kibis}}]{mand}
\bibinfo{author}{\bibfnamefont{S.}~\bibnamefont{Mandal}},
  \bibinfo{author}{\bibfnamefont{T.}~\bibnamefont{Liew}}, \bibnamefont{and}
  \bibinfo{author}{\bibfnamefont{O.}~\bibnamefont{Kibis}},
  \bibinfo{journal}{Physical Review A} \textbf{\bibinfo{volume}{97}},
  \bibinfo{pages}{043860} (\bibinfo{year}{2018}).

\bibitem[{\citenamefont{Morina et~al.}(2015)\citenamefont{Morina, Kibis,
  Pervishko, and Shelykh}}]{Mor2}
\bibinfo{author}{\bibfnamefont{S.}~\bibnamefont{Morina}},
  \bibinfo{author}{\bibfnamefont{O.~V.} \bibnamefont{Kibis}},
  \bibinfo{author}{\bibfnamefont{A.~A.} \bibnamefont{Pervishko}},
  \bibnamefont{and} \bibinfo{author}{\bibfnamefont{I.~A.}
  \bibnamefont{Shelykh}}, \bibinfo{journal}{Phys. Rev. B}
  \textbf{\bibinfo{volume}{91}}, \bibinfo{pages}{155312}
  (\bibinfo{year}{2015}).

\bibitem[{\citenamefont{Kibis}(2010)}]{KiMain}
\bibinfo{author}{\bibfnamefont{O.}~\bibnamefont{Kibis}},
  \bibinfo{journal}{Physical Review B} \textbf{\bibinfo{volume}{81}},
  \bibinfo{pages}{165433} (\bibinfo{year}{2010}).

\bibitem[{\citenamefont{Kibis et~al.}(2017)\citenamefont{Kibis, Dini, Iorsh,
  and Shelykh}}]{ki1210}
\bibinfo{author}{\bibfnamefont{O.}~\bibnamefont{Kibis}},
  \bibinfo{author}{\bibfnamefont{K.}~\bibnamefont{Dini}},
  \bibinfo{author}{\bibfnamefont{I.}~\bibnamefont{Iorsh}}, \bibnamefont{and}
  \bibinfo{author}{\bibfnamefont{I.}~\bibnamefont{Shelykh}},
  \bibinfo{journal}{Physical Review B} \textbf{\bibinfo{volume}{95}},
  \bibinfo{pages}{125401} (\bibinfo{year}{2017}).

\bibitem[{\citenamefont{Iurov et~al.}(2017{\natexlab{a}})\citenamefont{Iurov,
  Zhemchuzhna, Gumbs, and Huang}}]{ourJAP2017}
\bibinfo{author}{\bibfnamefont{A.}~\bibnamefont{Iurov}},
  \bibinfo{author}{\bibfnamefont{L.}~\bibnamefont{Zhemchuzhna}},
  \bibinfo{author}{\bibfnamefont{G.}~\bibnamefont{Gumbs}}, \bibnamefont{and}
  \bibinfo{author}{\bibfnamefont{D.}~\bibnamefont{Huang}},
  \bibinfo{journal}{Journal of Applied Physics} \textbf{\bibinfo{volume}{122}},
  \bibinfo{pages}{124301} (\bibinfo{year}{2017}{\natexlab{a}}).

\bibitem[{\citenamefont{Kibis et~al.}(2018)\citenamefont{Kibis, Dini, Iorsh,
  and Shelykh}}]{ki-spri}
\bibinfo{author}{\bibfnamefont{O.}~\bibnamefont{Kibis}},
  \bibinfo{author}{\bibfnamefont{K.}~\bibnamefont{Dini}},
  \bibinfo{author}{\bibfnamefont{I.}~\bibnamefont{Iorsh}}, \bibnamefont{and}
  \bibinfo{author}{\bibfnamefont{I.}~\bibnamefont{Shelykh}},
  \bibinfo{journal}{Semiconductors} \textbf{\bibinfo{volume}{52}},
  \bibinfo{pages}{523} (\bibinfo{year}{2018}).

\bibitem[{\citenamefont{Oka and Aoki}(2009)}]{oka}
\bibinfo{author}{\bibfnamefont{T.}~\bibnamefont{Oka}} \bibnamefont{and}
  \bibinfo{author}{\bibfnamefont{H.}~\bibnamefont{Aoki}},
  \bibinfo{journal}{Physical Review B} \textbf{\bibinfo{volume}{79}},
  \bibinfo{pages}{081406} (\bibinfo{year}{2009}).

\bibitem[{\citenamefont{D{\'o}ra et~al.}(2012)\citenamefont{D{\'o}ra, Cayssol,
  Simon, and Moessner}}]{dora}
\bibinfo{author}{\bibfnamefont{B.}~\bibnamefont{D{\'o}ra}},
  \bibinfo{author}{\bibfnamefont{J.}~\bibnamefont{Cayssol}},
  \bibinfo{author}{\bibfnamefont{F.}~\bibnamefont{Simon}}, \bibnamefont{and}
  \bibinfo{author}{\bibfnamefont{R.}~\bibnamefont{Moessner}},
  \bibinfo{journal}{Physical review letters} \textbf{\bibinfo{volume}{108}},
  \bibinfo{pages}{056602} (\bibinfo{year}{2012}).

\bibitem[{\citenamefont{Abergel and Chakraborty}(2009)}]{chak}
\bibinfo{author}{\bibfnamefont{D.}~\bibnamefont{Abergel}} \bibnamefont{and}
  \bibinfo{author}{\bibfnamefont{T.}~\bibnamefont{Chakraborty}},
  \bibinfo{journal}{Applied Physics Letters} \textbf{\bibinfo{volume}{95}},
  \bibinfo{pages}{062107} (\bibinfo{year}{2009}).

\bibitem[{\citenamefont{Calvo et~al.}(2018)\citenamefont{Calvo, Luna, Dal~Lago,
  and Torres}}]{to2}
\bibinfo{author}{\bibfnamefont{H.~L.} \bibnamefont{Calvo}},
  \bibinfo{author}{\bibfnamefont{J.~S.} \bibnamefont{Luna}},
  \bibinfo{author}{\bibfnamefont{V.}~\bibnamefont{Dal~Lago}}, \bibnamefont{and}
  \bibinfo{author}{\bibfnamefont{L.~E.~F.} \bibnamefont{Torres}},
  \bibinfo{journal}{Physical Review B} \textbf{\bibinfo{volume}{98}},
  \bibinfo{pages}{035423} (\bibinfo{year}{2018}).

\bibitem[{\citenamefont{Dal~Lago et~al.}(2017)\citenamefont{Dal~Lago, Morell,
  and Torres}}]{Tor1}
\bibinfo{author}{\bibfnamefont{V.}~\bibnamefont{Dal~Lago}},
  \bibinfo{author}{\bibfnamefont{E.~S.} \bibnamefont{Morell}},
  \bibnamefont{and} \bibinfo{author}{\bibfnamefont{L.~F.}
  \bibnamefont{Torres}}, \bibinfo{journal}{Physical Review B}
  \textbf{\bibinfo{volume}{96}}, \bibinfo{pages}{235409}
  (\bibinfo{year}{2017}).

\bibitem[{\citenamefont{Usaj et~al.}(2014)\citenamefont{Usaj, Perez-Piskunow,
  Torres, and Balseiro}}]{Tor2}
\bibinfo{author}{\bibfnamefont{G.}~\bibnamefont{Usaj}},
  \bibinfo{author}{\bibfnamefont{P.}~\bibnamefont{Perez-Piskunow}},
  \bibinfo{author}{\bibfnamefont{L.~F.} \bibnamefont{Torres}},
  \bibnamefont{and} \bibinfo{author}{\bibfnamefont{C.}~\bibnamefont{Balseiro}},
  \bibinfo{journal}{Physical Review B} \textbf{\bibinfo{volume}{90}},
  \bibinfo{pages}{115423} (\bibinfo{year}{2014}).

\bibitem[{\citenamefont{Iurov et~al.}(2013)\citenamefont{Iurov, Gumbs, Roslyak,
  and Huang}}]{ourTI}
\bibinfo{author}{\bibfnamefont{A.}~\bibnamefont{Iurov}},
  \bibinfo{author}{\bibfnamefont{G.}~\bibnamefont{Gumbs}},
  \bibinfo{author}{\bibfnamefont{O.}~\bibnamefont{Roslyak}}, \bibnamefont{and}
  \bibinfo{author}{\bibfnamefont{D.}~\bibnamefont{Huang}},
  \bibinfo{journal}{Journal of Physics: Condensed Matter}
  \textbf{\bibinfo{volume}{25}}, \bibinfo{pages}{135502}
  (\bibinfo{year}{2013}).

\bibitem[{\citenamefont{Yudin et~al.}(2016)\citenamefont{Yudin, Kibis, and
  Shelykh}}]{kiNJP}
\bibinfo{author}{\bibfnamefont{D.}~\bibnamefont{Yudin}},
  \bibinfo{author}{\bibfnamefont{O.}~\bibnamefont{Kibis}}, \bibnamefont{and}
  \bibinfo{author}{\bibfnamefont{I.}~\bibnamefont{Shelykh}},
  \bibinfo{journal}{New Journal of Physics} \textbf{\bibinfo{volume}{18}},
  \bibinfo{pages}{103014} (\bibinfo{year}{2016}).

\bibitem[{\citenamefont{Iorsh et~al.}(2017)\citenamefont{Iorsh, Dini, Kibis,
  and Shelykh}}]{ior1}
\bibinfo{author}{\bibfnamefont{I.}~\bibnamefont{Iorsh}},
  \bibinfo{author}{\bibfnamefont{K.}~\bibnamefont{Dini}},
  \bibinfo{author}{\bibfnamefont{O.}~\bibnamefont{Kibis}}, \bibnamefont{and}
  \bibinfo{author}{\bibfnamefont{I.}~\bibnamefont{Shelykh}},
  \bibinfo{journal}{Physical Review B} \textbf{\bibinfo{volume}{96}},
  \bibinfo{pages}{155432} (\bibinfo{year}{2017}).

\bibitem[{\citenamefont{Katsnelson et~al.}(2006)\citenamefont{Katsnelson,
  Novoselov, and Geim}}]{Kats}
\bibinfo{author}{\bibfnamefont{M.}~\bibnamefont{Katsnelson}},
  \bibinfo{author}{\bibfnamefont{K.}~\bibnamefont{Novoselov}},
  \bibnamefont{and} \bibinfo{author}{\bibfnamefont{A.}~\bibnamefont{Geim}},
  \bibinfo{journal}{Nature physics} \textbf{\bibinfo{volume}{2}},
  \bibinfo{pages}{620} (\bibinfo{year}{2006}).

\bibitem[{\citenamefont{Calvo et~al.}(2011)\citenamefont{Calvo, Pastawski,
  Roche, and Torres}}]{p2}
\bibinfo{author}{\bibfnamefont{H.~L.} \bibnamefont{Calvo}},
  \bibinfo{author}{\bibfnamefont{H.~M.} \bibnamefont{Pastawski}},
  \bibinfo{author}{\bibfnamefont{S.}~\bibnamefont{Roche}}, \bibnamefont{and}
  \bibinfo{author}{\bibfnamefont{L.~E.~F.} \bibnamefont{Torres}},
  \bibinfo{journal}{Applied Physics Letters} \textbf{\bibinfo{volume}{98}},
  \bibinfo{pages}{232103} (\bibinfo{year}{2011}).

\bibitem[{\citenamefont{Iurov et~al.}(2011)\citenamefont{Iurov, Gumbs, Roslyak,
  and Huang}}]{ouranomalous}
\bibinfo{author}{\bibfnamefont{A.}~\bibnamefont{Iurov}},
  \bibinfo{author}{\bibfnamefont{G.}~\bibnamefont{Gumbs}},
  \bibinfo{author}{\bibfnamefont{O.}~\bibnamefont{Roslyak}}, \bibnamefont{and}
  \bibinfo{author}{\bibfnamefont{D.}~\bibnamefont{Huang}},
  \bibinfo{journal}{Journal of Physics: Condensed Matter}
  \textbf{\bibinfo{volume}{24}}, \bibinfo{pages}{015303}
  (\bibinfo{year}{2011}).

\bibitem[{\citenamefont{Gumbs et~al.}(2014)\citenamefont{Gumbs, Iurov, Huang,
  Fekete, and Zhemchuzhna}}]{paula1}
\bibinfo{author}{\bibfnamefont{G.}~\bibnamefont{Gumbs}},
  \bibinfo{author}{\bibfnamefont{A.}~\bibnamefont{Iurov}},
  \bibinfo{author}{\bibfnamefont{D.}~\bibnamefont{Huang}},
  \bibinfo{author}{\bibfnamefont{P.}~\bibnamefont{Fekete}}, \bibnamefont{and}
  \bibinfo{author}{\bibfnamefont{L.}~\bibnamefont{Zhemchuzhna}}, in
  \emph{\bibinfo{booktitle}{AIP Conference Proceedings}}
  (\bibinfo{organization}{AIP}, \bibinfo{year}{2014}), vol.
  \bibinfo{volume}{1590}, pp. \bibinfo{pages}{134--142}.

\bibitem[{\citenamefont{Morina et~al.}(2018)\citenamefont{Morina, Dini, Iorsh,
  and Shelykh}}]{Mor1}
\bibinfo{author}{\bibfnamefont{S.}~\bibnamefont{Morina}},
  \bibinfo{author}{\bibfnamefont{K.}~\bibnamefont{Dini}},
  \bibinfo{author}{\bibfnamefont{I.~V.} \bibnamefont{Iorsh}}, \bibnamefont{and}
  \bibinfo{author}{\bibfnamefont{I.~A.} \bibnamefont{Shelykh}},
  \bibinfo{journal}{ACS Photonics} \textbf{\bibinfo{volume}{5}},
  \bibinfo{pages}{1171} (\bibinfo{year}{2018}).

\bibitem[{\citenamefont{Iurov et~al.}(2017{\natexlab{b}})\citenamefont{Iurov,
  Gumbs, and Huang}}]{ourPQE}
\bibinfo{author}{\bibfnamefont{A.}~\bibnamefont{Iurov}},
  \bibinfo{author}{\bibfnamefont{G.}~\bibnamefont{Gumbs}}, \bibnamefont{and}
  \bibinfo{author}{\bibfnamefont{D.}~\bibnamefont{Huang}},
  \bibinfo{journal}{Journal of Modern Optics} \textbf{\bibinfo{volume}{64}},
  \bibinfo{pages}{913} (\bibinfo{year}{2017}{\natexlab{b}}).

\bibitem[{\citenamefont{Kristinsson et~al.}(2016)\citenamefont{Kristinsson,
  Kibis, Morina, and Shelykh}}]{kisrep}
\bibinfo{author}{\bibfnamefont{K.}~\bibnamefont{Kristinsson}},
  \bibinfo{author}{\bibfnamefont{O.}~\bibnamefont{Kibis}},
  \bibinfo{author}{\bibfnamefont{S.}~\bibnamefont{Morina}}, \bibnamefont{and}
  \bibinfo{author}{\bibfnamefont{I.}~\bibnamefont{Shelykh}},
  \bibinfo{journal}{Scientific reports} \textbf{\bibinfo{volume}{6}},
  \bibinfo{pages}{20082} (\bibinfo{year}{2016}).

\bibitem[{\citenamefont{Li et~al.}(2017)\citenamefont{Li, Cao, Wu, and
  Louie}}]{liklein}
\bibinfo{author}{\bibfnamefont{Z.}~\bibnamefont{Li}},
  \bibinfo{author}{\bibfnamefont{T.}~\bibnamefont{Cao}},
  \bibinfo{author}{\bibfnamefont{M.}~\bibnamefont{Wu}}, \bibnamefont{and}
  \bibinfo{author}{\bibfnamefont{S.~G.} \bibnamefont{Louie}},
  \bibinfo{journal}{Nano letters} \textbf{\bibinfo{volume}{17}},
  \bibinfo{pages}{2280} (\bibinfo{year}{2017}).

\bibitem[{\citenamefont{Illes and Nicol}(2017)}]{akl}
\bibinfo{author}{\bibfnamefont{E.}~\bibnamefont{Illes}} \bibnamefont{and}
  \bibinfo{author}{\bibfnamefont{E.}~\bibnamefont{Nicol}},
  \bibinfo{journal}{Physical Review B} \textbf{\bibinfo{volume}{95}},
  \bibinfo{pages}{235432} (\bibinfo{year}{2017}).

\bibitem[{\citenamefont{Urban et~al.}(2011)\citenamefont{Urban, Bercioux,
  Wimmer, and H{\"a}usler}}]{dkl}
\bibinfo{author}{\bibfnamefont{D.~F.} \bibnamefont{Urban}},
  \bibinfo{author}{\bibfnamefont{D.}~\bibnamefont{Bercioux}},
  \bibinfo{author}{\bibfnamefont{M.}~\bibnamefont{Wimmer}}, \bibnamefont{and}
  \bibinfo{author}{\bibfnamefont{W.}~\bibnamefont{H{\"a}usler}},
  \bibinfo{journal}{Physical Review B} \textbf{\bibinfo{volume}{84}},
  \bibinfo{pages}{115136} (\bibinfo{year}{2011}).

\bibitem[{\citenamefont{Illes et~al.}(2015)\citenamefont{Illes, Carbotte, and
  Nicol}}]{illes01}
\bibinfo{author}{\bibfnamefont{E.}~\bibnamefont{Illes}},
  \bibinfo{author}{\bibfnamefont{J.}~\bibnamefont{Carbotte}}, \bibnamefont{and}
  \bibinfo{author}{\bibfnamefont{E.}~\bibnamefont{Nicol}},
  \bibinfo{journal}{Physical Review B} \textbf{\bibinfo{volume}{92}},
  \bibinfo{pages}{245410} (\bibinfo{year}{2015}).

\bibitem[{\citenamefont{Raoux et~al.}(2014)\citenamefont{Raoux, Morigi, Fuchs,
  Pi{\'e}chon, and Montambaux}}]{piech1}
\bibinfo{author}{\bibfnamefont{A.}~\bibnamefont{Raoux}},
  \bibinfo{author}{\bibfnamefont{M.}~\bibnamefont{Morigi}},
  \bibinfo{author}{\bibfnamefont{J.-N.} \bibnamefont{Fuchs}},
  \bibinfo{author}{\bibfnamefont{F.}~\bibnamefont{Pi{\'e}chon}},
  \bibnamefont{and}
  \bibinfo{author}{\bibfnamefont{G.}~\bibnamefont{Montambaux}},
  \bibinfo{journal}{Physical review letters} \textbf{\bibinfo{volume}{112}},
  \bibinfo{pages}{026402} (\bibinfo{year}{2014}).

\bibitem[{\citenamefont{Pi{\'e}chon et~al.}(2015)\citenamefont{Pi{\'e}chon,
  Fuchs, Raoux, and Montambaux}}]{piech2}
\bibinfo{author}{\bibfnamefont{F.}~\bibnamefont{Pi{\'e}chon}},
  \bibinfo{author}{\bibfnamefont{J.}~\bibnamefont{Fuchs}},
  \bibinfo{author}{\bibfnamefont{A.}~\bibnamefont{Raoux}}, \bibnamefont{and}
  \bibinfo{author}{\bibfnamefont{G.}~\bibnamefont{Montambaux}}, in
  \emph{\bibinfo{booktitle}{Journal of Physics: Conference Series}}
  (\bibinfo{organization}{IOP Publishing}, \bibinfo{year}{2015}), vol.
  \bibinfo{volume}{603}, p. \bibinfo{pages}{012001}.

\bibitem[{\citenamefont{Illes and Nicol}(2016)}]{illes02}
\bibinfo{author}{\bibfnamefont{E.}~\bibnamefont{Illes}} \bibnamefont{and}
  \bibinfo{author}{\bibfnamefont{E.}~\bibnamefont{Nicol}},
  \bibinfo{journal}{Physical Review B} \textbf{\bibinfo{volume}{94}},
  \bibinfo{pages}{125435} (\bibinfo{year}{2016}).

\bibitem[{\citenamefont{Biswas and Ghosh}(2016)}]{tutul1}
\bibinfo{author}{\bibfnamefont{T.}~\bibnamefont{Biswas}} \bibnamefont{and}
  \bibinfo{author}{\bibfnamefont{T.~K.} \bibnamefont{Ghosh}},
  \bibinfo{journal}{Journal of Physics: Condensed Matter}
  \textbf{\bibinfo{volume}{28}}, \bibinfo{pages}{495302}
  (\bibinfo{year}{2016}).

\bibitem[{\citenamefont{Biswas and Ghosh}(2018)}]{tutul2}
\bibinfo{author}{\bibfnamefont{T.}~\bibnamefont{Biswas}} \bibnamefont{and}
  \bibinfo{author}{\bibfnamefont{T.~K.} \bibnamefont{Ghosh}},
  \bibinfo{journal}{Journal of Physics: Condensed Matter}
  \textbf{\bibinfo{volume}{30}}, \bibinfo{pages}{075301}
  (\bibinfo{year}{2018}).

\bibitem[{\citenamefont{Tabert and Nicol}(2014)}]{SilMain}
\bibinfo{author}{\bibfnamefont{C.~J.} \bibnamefont{Tabert}} \bibnamefont{and}
  \bibinfo{author}{\bibfnamefont{E.~J.} \bibnamefont{Nicol}},
  \bibinfo{journal}{Physical Review B} \textbf{\bibinfo{volume}{89}},
  \bibinfo{pages}{195410} (\bibinfo{year}{2014}).

\bibitem[{\citenamefont{Ezawa}(2012)}]{EzawaSi}
\bibinfo{author}{\bibfnamefont{M.}~\bibnamefont{Ezawa}}, \bibinfo{journal}{New
  Journal of Physics} \textbf{\bibinfo{volume}{14}}, \bibinfo{pages}{033003}
  (\bibinfo{year}{2012}).

\bibitem[{\citenamefont{Iurov et~al.}(2018)\citenamefont{Iurov, Gumbs, and
  Huang}}]{ournewprb}
\bibinfo{author}{\bibfnamefont{A.}~\bibnamefont{Iurov}},
  \bibinfo{author}{\bibfnamefont{G.}~\bibnamefont{Gumbs}}, \bibnamefont{and}
  \bibinfo{author}{\bibfnamefont{D.}~\bibnamefont{Huang}},
  \bibinfo{journal}{Physical Review B} \textbf{\bibinfo{volume}{98}},
  \bibinfo{pages}{075414} (\bibinfo{year}{2018}).

\bibitem[{\citenamefont{Xu et~al.}(2017)\citenamefont{Xu, Huang, Huang, and
  Lai}}]{hua10}
\bibinfo{author}{\bibfnamefont{H.-Y.} \bibnamefont{Xu}},
  \bibinfo{author}{\bibfnamefont{L.}~\bibnamefont{Huang}},
  \bibinfo{author}{\bibfnamefont{D.}~\bibnamefont{Huang}}, \bibnamefont{and}
  \bibinfo{author}{\bibfnamefont{Y.-C.} \bibnamefont{Lai}},
  \bibinfo{journal}{Physical Review B} \textbf{\bibinfo{volume}{96}},
  \bibinfo{pages}{045412} (\bibinfo{year}{2017}).

\bibitem[{\citenamefont{Berry}(1984)}]{A127}
\bibinfo{author}{\bibfnamefont{M.~V.} \bibnamefont{Berry}},
  \bibinfo{journal}{Proc. R. Soc. Lond. A} \textbf{\bibinfo{volume}{392}},
  \bibinfo{pages}{45} (\bibinfo{year}{1984}).

\bibitem[{\citenamefont{Xiao et~al.}(2010)\citenamefont{Xiao, Chang, and
  Niu}}]{A129}
\bibinfo{author}{\bibfnamefont{D.}~\bibnamefont{Xiao}},
  \bibinfo{author}{\bibfnamefont{M.-C.} \bibnamefont{Chang}}, \bibnamefont{and}
  \bibinfo{author}{\bibfnamefont{Q.}~\bibnamefont{Niu}},
  \bibinfo{journal}{Reviews of modern physics} \textbf{\bibinfo{volume}{82}},
  \bibinfo{pages}{1959} (\bibinfo{year}{2010}).

\bibitem[{\citenamefont{Illes}(2017)}]{A1Thesis}
\bibinfo{author}{\bibfnamefont{E.}~\bibnamefont{Illes}}, Ph.D. thesis
  (\bibinfo{year}{2017}),
  \urlprefix\url{http://atrium.lib.uoguelph.ca/xmlui/handle/10214/11512}.

\bibitem[{\citenamefont{Louvet et~al.}(2015)\citenamefont{Louvet, Delplace,
  Fedorenko, and Carpentier}}]{A131}
\bibinfo{author}{\bibfnamefont{T.}~\bibnamefont{Louvet}},
  \bibinfo{author}{\bibfnamefont{P.}~\bibnamefont{Delplace}},
  \bibinfo{author}{\bibfnamefont{A.~A.} \bibnamefont{Fedorenko}},
  \bibnamefont{and}
  \bibinfo{author}{\bibfnamefont{D.}~\bibnamefont{Carpentier}},
  \bibinfo{journal}{Physical Review B} \textbf{\bibinfo{volume}{92}},
  \bibinfo{pages}{155116} (\bibinfo{year}{2015}).

\bibitem[{\citenamefont{Aharonov and Bohm}(1959)}]{A1AB}
\bibinfo{author}{\bibfnamefont{Y.}~\bibnamefont{Aharonov}} \bibnamefont{and}
  \bibinfo{author}{\bibfnamefont{D.}~\bibnamefont{Bohm}},
  \bibinfo{journal}{Physical Review} \textbf{\bibinfo{volume}{115}},
  \bibinfo{pages}{485} (\bibinfo{year}{1959}).

\bibitem[{\citenamefont{Arfken and Weber}(1999)}]{arfken}
\bibinfo{author}{\bibfnamefont{G.~B.} \bibnamefont{Arfken}} \bibnamefont{and}
  \bibinfo{author}{\bibfnamefont{H.~J.} \bibnamefont{Weber}},
  \emph{\bibinfo{title}{Mathematical methods for physicists}}
  (\bibinfo{year}{1999}).

\end{thebibliography}

\end{document}